\documentclass[fleqn,usenatbib]{mnras}

\usepackage{newtxtext,newtxmath}
\usepackage[bottom]{footmisc}

\usepackage[T1]{fontenc}

\DeclareRobustCommand{\VAN}[3]{#2}
\let\VANthebibliography\thebibliography
\def\thebibliography{\DeclareRobustCommand{\VAN}[3]{##3}\VANthebibliography}

\usepackage{graphicx}	
\usepackage{amsmath}	
\newcommand{\lephare}{{\tt{LePhare}}}
\newcommand{\eazy}{{\tt{EAZY}}}

\newcommand\cigale{{\tt cigale}}
\newcommand\fagn{{\tt frac$_{\rm AGN}$}}
\newcommand\mkms{\rm km \, s^{-1}}

\title[MIRI-selected galaxies in SMACS0723]
{EPOCHS VIII. An Insight into MIRI-selected Galaxies in SMACS-0723 and the Benefits of Deep MIRI Photometry in Revealing AGN and the Dusty Universe}

\author[Li et al.]{
Qiong Li,$^{1}$\thanks{E-mail: qiong.li@manchester.ac.uk}
Christopher J. Conselice,$^{1}$
Nathan Adams,$^{1}$
James A.\@ A.\@ Trussler,$^{1}$
Duncan Austin,$^{1}$
\newauthor
Thomas Harvey,$^{1}$
Leonardo Ferreira,$^{4}$
Joseph Caruana,$^{2,3}$
Katherine Ormerod,$^{1}$
Ignas Juodžbalis$^{5,6}$
\\
$^{1}$ Jodrell Bank Centre for Astrophysics, University of Manchester, Oxford Road, Manchester UK \\
$^{2}$ Department of Physics, University of Malta, Msida MSD 2080, Malta \\
$^{3}$ Institute of Space Sciences \& Astronomy, University of Malta, Msida MSD 2080, Malta \\
$^{4}$ School of Physics and Astronomy, University of Victoria, Victoria, BC, Canada \\
$^{5}$ Kavli Institute for Cosmology, University of Cambridge, Madingley Road, Cambridge, CB3 0HA, UK \\
$^{6}$ Cavendish Laboratory, University of Cambridge, 19 JJ Thomson Avenue, Cambridge CB3 0HE, UK} 

\date{Accepted XXX. Received YYY; in original form ZZZ}

\pubyear{2023}

\begin{document}
\label{firstpage}
\pagerange{\pageref{firstpage}--\pageref{lastpage}}
\maketitle

\begin{abstract}
We present the analysis of the stellar population and star formation history of 181 MIRI selected galaxies at redshift $0-3.5$ in the massive galaxy cluster field SMACS J0723.3–7327, commonly referred to as SMACS0723, using the James Webb Space Telescope (JWST) Mid-Infrared Instrument (MIRI). We combine the data with the JWST Near Infrared Camera (NIRCam) catalogue, in conjunction with the Hubble Space Telescope (HST) WFC3/IR and ACS imaging. We find that the MIRI bands capture PAH features and dust emission, significantly enhancing the accuracy of photometric redshift and measurements of the physical properties of these galaxies. The median photo-$z$'s of galaxies with MIRI data are found to have a small 0.1\% difference from spectroscopic redshifts and reducing the error by 20\%. With MIRI data included in SED fits,  we find that the measured stellar masses are unchanged, while the star formation rate is systematically lower by 0.1 dex. We also fit the median SED of active galactic nuclei (AGN) and star forming galaxies (SFG) separately. MIRI data provides tighter constraints on the AGN contribution, reducing the typical AGN contributions by $\sim$14\%. In addition, we also compare the median SED obtained with and without MIRI, and we find that including MIRI data yields steeper optical and UV slopes, indicating bluer colours, lower dust attenuation, and younger stellar populations. In the future, MIRI/MRS will enhance our understanding by providing more detailed spectral information and allowing for the study of specific emission features and diagnostics associated with AGN.
\end{abstract}




\section{Introduction} \label{sec:intro}


In the vast expanse of the universe, many galaxies remain hidden behind a veil of dust, rendering them challenging to observe using traditional optical telescopes (e.g. \citealt{Asboth2016,Fudamoto2017,Reuter2020}). Dust particles can absorb or scatter the emitted light, obstructing our view and limiting our understanding of their properties and evolution. However, the advent of the James Webb Space Telescope (JWST) and its successful commissioning has opened up a new era of exploration at infrared wavelengths \citep{Menzel2023,Rigby2023}.

JWST has started to revolutionise our ability to study the dusty universe by enabling deep imaging and spectroscopy in the 1-30 $\mu$m wavelength range. Its new capabilities, including high sensitivity and exceptional spatial resolution, have propelled our investigations into the basic features of galaxies (e.g. \citealt{Pontoppidan2022,Adams2023,Castellano2022, Naidu2022,Yan2022,Harikane2022}). By delving deep into the universe with this imaging, we can uncover intricate details about galaxy structures, stellar populations, and the interplay between stars, gas, and dust. Furthermore, the JWST's infrared observations provide valuable insights into star formation processes, dust distribution, and the activity of supermassive black holes at the centers of galaxies.

\begin{figure*}
\centering
 \includegraphics[width=0.98\textwidth]{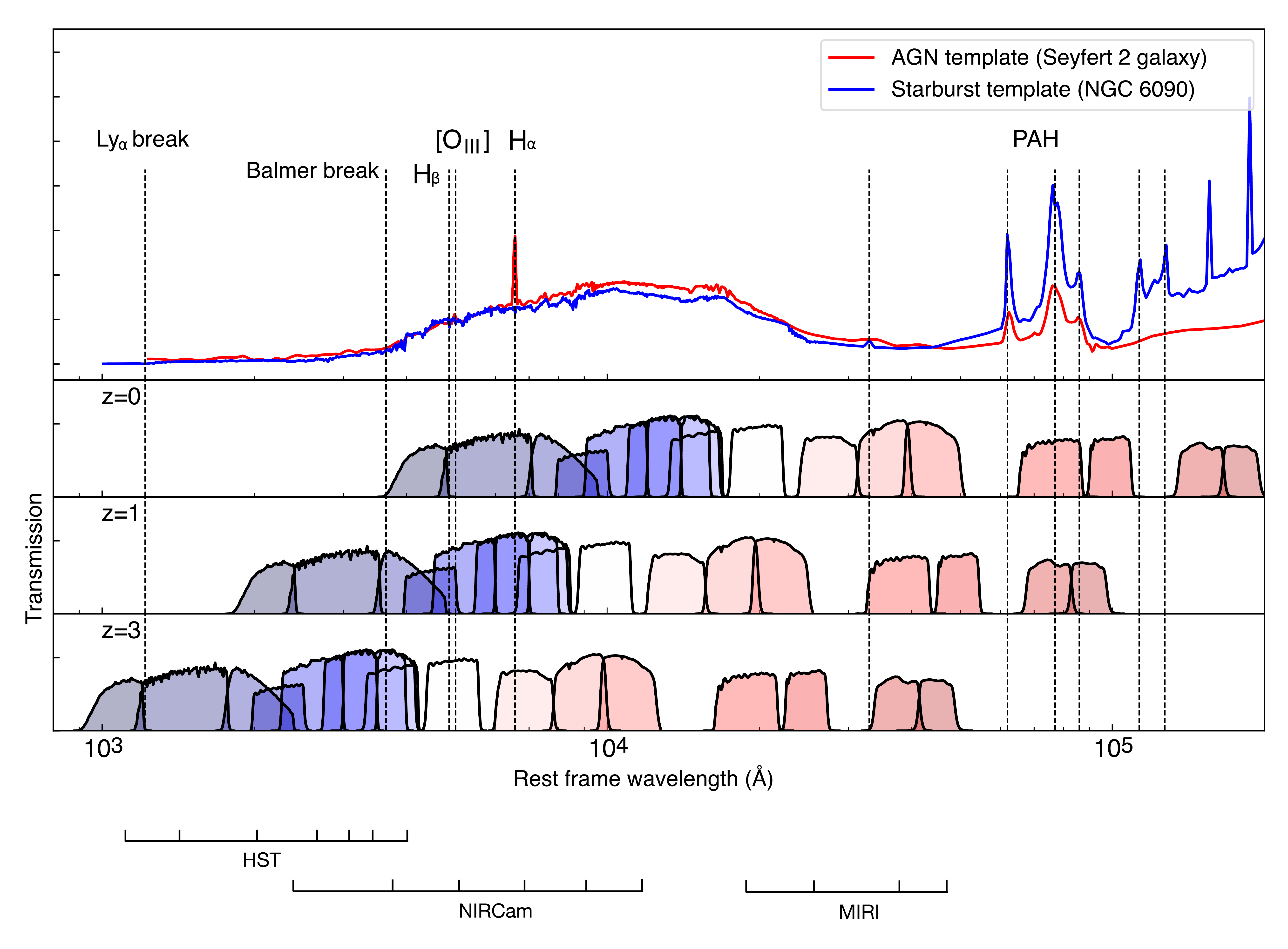}
    \caption{Plot showing the JWST and HST filters we use as well as SEDs for representative AGNs and SFGs. The broadband coverage of the AGN (Seyfert 2 galaxy) and starburst galaxy (NGC6090) templates ($\lambda F_{\lambda}$, in relative units of erg s$^{-1}$) at different redshift bins \citep{Weedman2006} are shown. The top panel presents the AGN and star forming galaxy templates, while the bottom panel displays the relative transmission functions for various filters: HST/ACS and WCS3/IR (F435W, F606W, F814W, F105W, F125W, F140W, and F160W), JWST/NIRCam (F090W, F150W, F200W, F277W, F356W, and F444W), and JWST/MIRI (F770W, F1000W, F1500W, and F1800W). Emission lines and PAH features are appropriately labelled. Notably, the MIRI data enable us to probe the spectral energy distributions of galaxies up to $\sim5\mu$m (at $z=3$) in the rest-frame, facilitating the characterization of PAH features and dust emission.}
    \label{fig:filter}
\end{figure*}

The emission from dust can be divided into three broad components as wavelength increases towards the red. Around rest-frame 8 $\mu$m, the mid-infrared range is dominated by features known as polycyclic aromatic hydrocarbon (PAH) bands \citep{Allamandola1989}. These PAHs can absorb UV photons and re-emit the absorbed energy as fluorescence at longer wavelengths, typically in the mid-infrared range. As the wavelength increases beyond the mid-infrared range, the emission is progressively taken over by very small, warm grains. At higher radiation field intensities, equilibrium emission from these warm grains becomes dominant. Beyond 100 $\mu$m, the emission is increasingly attributed to larger, relatively cold grains.

While the {\it Spitzer} Space Telescope allowed observations in this mid-infrared range, it had severe limitations in sensitivity and resolution at longer wavelengths (e.g., \citealt{Ashby2015,Timlin2016,Nayyeri2018}). The James Webb Space Telescope's Mid-Infrared Instrument (MIRI) has made significant advancements over this, offering higher sensitivity at a magnitude limit as deep as $\sim$29 mag (and perhaps beyond) and with sub-arcsecond resolution \citep{Wright2023,Rigby2023}. 
The advanced capabilities of MIRI thus enable more precise investigations into the impact of dust on star formation and galaxy evolution, as well as the analysis of PAH features in the mid-infrared (see Figure~\ref{fig:filter}), surpassing the limitations of optical and earlier infrared observations. In principle longer wavelengths can be used to find AGN and this is another advantage that MIRI has over what can be carried out with just NIRCam to find and characterise these objects, although see \citet[][]{Juodvzbalis2023}.

With these motivations in mind, we have selected a well-studied, strong lensing galaxy cluster field, SMACS\,0723 \citep{2007ApJ...663..717M,Ebeling2010,Repp2018} to carry out an analysis of the uses of MIRI data for uncovering galaxy properties. Previous research on this cluster field has been conducted using various telescopes and instruments, including {\it Chandra, VLT/MUSE,} {\it Subaru}, the {\it Hubble Space Telescope} (Reionization Lensing Cluster Survey; \citealt{Coe2019}), and {\it Planck} (e.g., \citealt{Richard2021,Lagattuta2022,Golubchik2022,Mahler2023}). \citet{Mahler2023} determined the cluster redshift to be $z=0.3877$ based on a sample of 26 spectroscopically confirmed cluster members. They also derived a cluster velocity dispersion of $\sigma \sim 1180\pm160$ km s$^{-1}$. According to the Planck estimation, the total mass of the cluster is approximately 8.39$\times 10^{14}$M$_\odot$ \citep{Coe2019}. Previous infrared observations with the {\it Spitzer} and {\it Herschel} Space Telescopes have revealed the presence of a population of dusty, infrared-luminous, red-sequence galaxies in the SMACS0723 field.

In this paper, we use JWST MIRI observations of SMACS0723 to study the role of MIRI in measuring photometric redshifts of distant galaxies and to study the physical properties of the potentially dusty and AGN galaxies which are obscured at optical bands. This is important as we know that the fraction and amount of AGN at high-$z$ is perhaps surprisingly high \citep[e.g.,][]{Juodvzbalis2023}.  Thus it is important to determine how we can measure the amount of AGN and their contribution to galaxy SEDs.  Thus, this paper focuses on the selection and analysis of dusty galaxies selected by MIRI bands in conjunction with HST and JWST/NIRCam data. 

\begin{figure*}
\centering
 \includegraphics[width=0.65\textwidth]{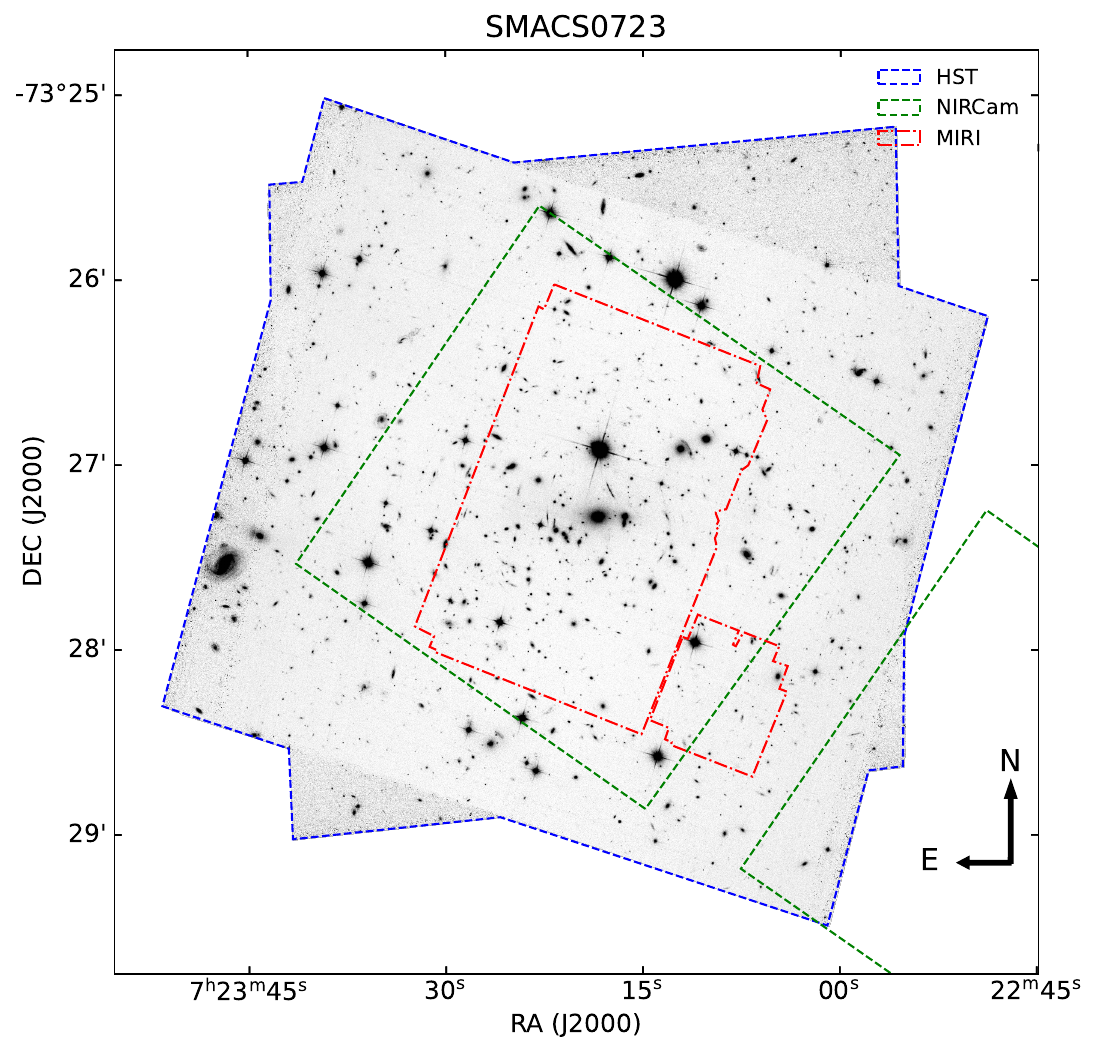}
    \caption{The SMACS0723 fields of view overlaid on HST images (R:JWST/MIRI, G:JWST/NIRCam, B:HST). Before generating the catalog, 
    we produce a mask to avoid the diffraction spikes of bright stars and image artifacts. These masks cover diffraction spikes, the few remaining snowballs, regions of intra-cluster medium, and a buffer around the edges of the images.  The imaging data is from HST, the green dotted boxes show the coverage of NIRCam, and the red dashed lines show the area imaged by MIRI.}
    \label{fig:fov}
\end{figure*}

The structure of the paper is organised as follows. We describe the JWST and the ancillary datasets used in this study and the data reduction process in \S\ref{sec:obs}. We also describe the catalog generation process. In \S\ref{sec:result}, we present the MIRI selected sample and the physical properties from the spectral energy distribution (SED) fitting for the galaxy. In \S\ref{sec:discussion}, our study focuses on the notable advancements achieved through the utilisation of MIRI data. We examine the enhancements it brings to various aspects, such as the accuracy of redshift measurements, the characterisation of star populations in galaxies, and the impact on the SED analysis of both active galactic nuclei (AGN) and star-forming galaxies (SFG). In \S\ref{sec:Conclusion}, we provide a comprehensive summary of our findings and discuss the potential avenues for future research in this field.

Throughout this paper, we assume a flat cosmological model with $\Omega_{\Lambda} = 0.7, \Omega_{m} = 0.3$ and $H_0 = 70 \mkms \, \rm Mpc^{-1}$. All magnitudes used in this paper are in the AB system \citep{Oke1983}.

\section{Data Reductions and Catalog}\label{sec:obs}

\subsection{JWST NIRCam observations}

Observations of the SMACS-0723 galaxy cluster were taken on 2022 June 06, as part of the Early Release Observations (ERO) programme (ID: 2736, PI: K. Pontoppidan, \citealt{Pontoppidan2022}). The observations consist of 6 NIRCam photometric bands F090W, F150W, F200W, F277W, F356W, and F444W. The total integration time is 12.5 hr. Our NIRCam image reduction is performed using the procedure of \citet{Ferreira2022} and \citet{Adams2023}. Below we summarise the procedure. The data were processed using the JWST Calibration Pipeline (v1.8.2 and CRDS v0995) using the default parameters for stages 1 and 2.   This was the most up-to-date version at the time of writing, and includes the second round of post-flight calibrations. We then apply the 1/f noise correction\footnote{\url{https://github.com/chriswillott/jwst}} derived by Chris Willott after stage 2. After stage 3, we subtract an initial flat background and carry out a 2-dimensional background subtraction. Then we align the final F444W image onto a GAIA-derived WCS using \texttt{tweakreg}, as  part of the DrizzlePac python package. We then match all remaining filters to this derived F444W WCS.\footnote{\url{https://github.com/spacetelescope/drizzlepac}} We then pixel-match the images to the F444W image with the use of \texttt{astropy reproject}.\footnote{\url{https://reproject.readthedocs.io/en/stable/}} The final resolution of the drizzled images is 0.03 arcseconds/pixel. 

We use the {\tt SExtractor} \citep{Bertin1996} version 2.8.6 to identify our sources. We run this in dual-image mode with the F444W image used for object selection. Here the apertures of all measurements should be consistent. MIRI's PSF FWHM is 0.5 arcseconds in the F1500W filter. Thus we conduct forced circular aperture photometry within 1.0 arcsecond diameters. We perform the aperture correction derived from simulated \texttt{WebbPSF} point spread functions\footnote{\url{https://jwst-docs.stsci.edu/jwst-near-infrared-camera/nircam-performance/nircam-point-spread-functions}} for each NIRCam band. We experimented with many different aperture photometry measurement methods and found that this one is the best for recovering accurately the fluxes of our galaxies.  The effects of galactic extinction are negligible in these IR bands ($<0.1$ mag), and thus are not applied.

\subsection{JWST MIRI observations}

MIRI observations for this field were taken on June 14th 2022, covering a specific area measuring 112\arcsec6$\times$73\arcsec5. The data acquisition included observations in the F770W, F1000W, F1500W, and F1800W bands within this field. Two versions of the reduced data are generated for analysis. In the first version, the data is processed using the {\tt grizli} reduced by Brammer et al. in prep\footnote{Images and catalogs of JWST/MIRI in the SMACS0723 field processed with the {\tt grizli} software pipeline: \url{https://zenodo.org/record/6874301}}. For the second version, MIRI images are acquired from the Mikulski Archive for Space Telescopes (MAST), and the data underwent reduction using the standard JWST pipeline, similar to the process utilised for NIRCam data. A comparative analysis of the standard JWST pipeline reduced images reveal the presence of pronounced background patterns, specifically stripes and gradients, predominantly around the edges of the images. However, the central region of the image exhibits no discernible impact from these artefacts. Consequently, in this paper the {\tt grizli} reduced images are employed due to their superior quality within the central region. The resulting drizzled images have a resolution of 0.04\arcsec.

We then align the images to NIRCam F444W matching systems with separations $\Delta<0.05$\arcsec\,. We then we run {\tt SExtractor} version 2.8.6 \citep{Bertin1996} in dual-image mode to detect objects in each field. 
The detection image we use is MIRI F770W. We use the F770W filter as it has the best sensitivity and angular resolution in the MIRI bands. The apertures of 1.0 arcsecs are the same as before. We also perform aperture corrections derived from simulated \texttt{WebbPSF} MIRI point spread functions\footnote{\url{https://jwst-docs.stsci.edu/jwst-mid-infrared-instrument/miri-performance/miri-point-spread-functions}} for each band.  This aperture corrections are essential as it allows us to measure photometry on different bands and then to normalise these measurements by correcting for the effects of using an aperture which by its nature limits the amount of flux measured.

\subsection{HST imaging observations}

HST observations of SMACS0723 are from the Reionization Lensing Cluster Survey (RELICS). This survey observed 41 massive galaxy clusters with Hubble and Spitzer at 0.4–1.7$\mu$m and 3.0–5.0$\mu$m, respectively. SMACS0723 (ID: GO 14017; \citealt{Coe2019}) was observed in one WFC3/IR pointing, with a total of ten orbits in WCS3/IR (F105W, F125W, F140W, and F160W) and ACS imaging (F435W, F606W, and F814W). The observational details and the HST data reduction are available from \citet{Coe2019}.  The image resolution is 0.06\arcsec\,. 

As mentioned before, before the source extraction we align the HST images to NIRCam F444W to a level of $\Delta<0.05$\arcsec\. Then we run  {\tt SExtractor} version 2.8.6 \citep{Bertin1996} in dual-image mode to detect objects in the field with an aperture of 1.0 arcsec for photometry measured in each filter image. The weighted stack of all the HST images is the input detection image, the same as that in \citet{Coe2019}. 

We also perform aperture corrections based on the ACS/WFC\footnote{\url{https://www.stsci.edu/hst/instrumentation/acs/data-analysis/aperture-corrections}} and WFC3/IR PSF\footnote{\url{https://www.stsci.edu/hst/instrumentation/wfc3/data-analysis/photometric-calibration/ir-encircled-energy}} encircled energy fraction.
We correct all photometry for Galactic extinction using the IR dust emission maps of \citet{Schlafly2011}.

\subsection{Source Photometry and Cataloguing}\label{sec:Catalog}

To generate a matched catalog for all the sources in SMACS0723, we use {\tt TOPCAT} to combine SExtractor's HST and JWST catalogs. The maximum separation allowed is 0.3\arcsec, which is a good compromise between the false positive rate achieved and how restricted it is due to the size of MIRI's PSF.  For the final catalogue, we use a forced circular 1\arcsec\, diameter aperture. This diameter is chosen to enclose the central/brightest $\sim$86 per cent of the flux of a point source of NIRCam and $\sim$83 per cent of MIRI, enabling us to use the highest SNR pixels to calculate galaxy colours while avoiding reliance on strong aperture corrections that can be as high as the actual measurements made. It is also consistent with the circular apertures of 0.9\arcsec\, diameter in \citet{Papovich2023}. Additionally, we create a composite mask to avoid image artefacts. These masks cover diffraction spikes, the few remaining snowballs in the NIRCam imaging, as well as regions of intra-cluster medium (in the NIRCam modules containing any foreground cluster), and a buffer around the edges of the observations. The remaining total unmasked region is $\sim2.3$ arcmin$^2$.  We plot the NIRCam and MIRI observations overlaid on the HST ACS F606W image, in Figure~\ref{fig:fov}.

{\tt SExtractor} is known to underestimate the photometric errors of sources. To ensure accurate measurements, we calculate the local depth of our final images. We place circular apertures (1\arcsec\,) in empty regions that are at least 1 arcsecond away from real sources in our images. We use the measured background flux in these apertures to derive a median depth for each field. Finally, we calculate the photometric errors for each individual source using the nearest 200 empty apertures to determine the local depth. The 5 sigma depths of each band can be found in Table~\ref{tab:corr_depth}.

Finally, we use robust methods to construct the final samples. The relevant selection criteria are described in Section~\ref{sec:MIRI_select}. In total, 181 galaxies are matched and meet our selection criteria. To comprehensively detect all sources in this field, especially at high redshift, we use NIRCam as the detection image and strive to determine the corresponding measurements for HST and MIRI. Specifically, we employ {\tt SExtractor++} \citep{Bertin2022} in dual-image mode to measure HST and MIRI fluxes for the NIRCam detections. For high-redshift galaxies at $z>6.5$, their blue-ward bands (HST) are anticipated to appear faint or undetected due to the Lyman break. Out of the total of 12 candidates at $z>6.5$ identified by NIRCam, unfortunately, these candidates are not within the coverage of MIRI and HST. More detailed analysis of this $z>6.5$ sample can be found in our EPOCHS paper I (Conselice in preparation).

\begin{figure}
    \includegraphics[width=8.5cm]{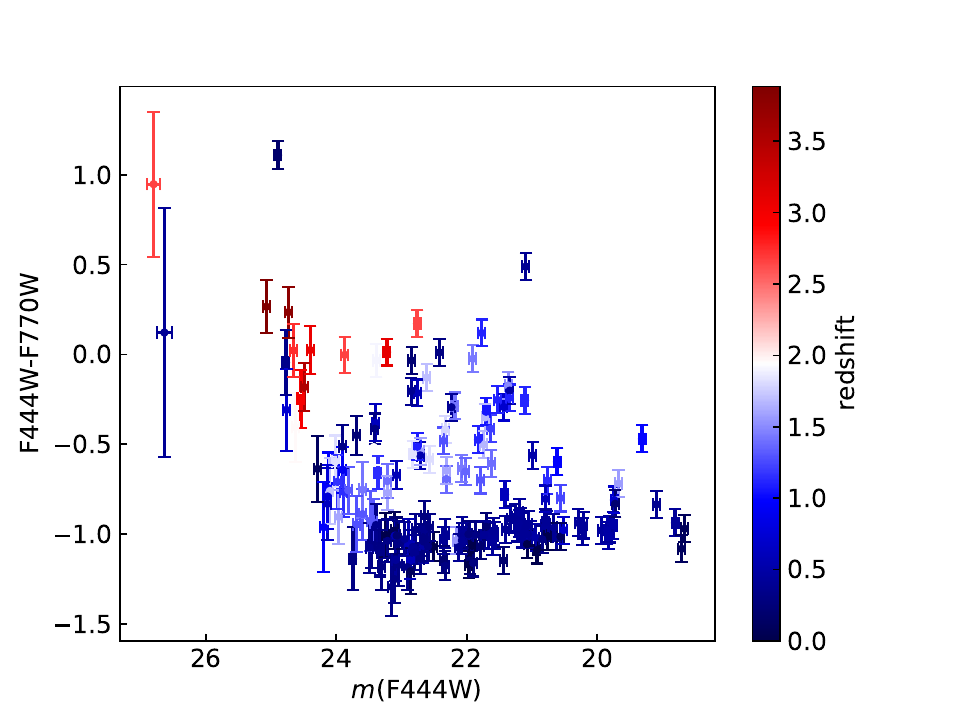}
    \caption{Plot of observed NIRCam and MIRI mag-colour diagram for the matched robust galaxies in SMACS0723 field. The magnitude error is calculated using measurements of the local depth. The redder colour corresponds to higher redshift galaxies. A gradient in redshift can clearly be seen in the F444W-F770W colour.}
    \label{fig:mag_colors}
\end{figure}

\begin{figure}
 \includegraphics[width=0.45\textwidth]{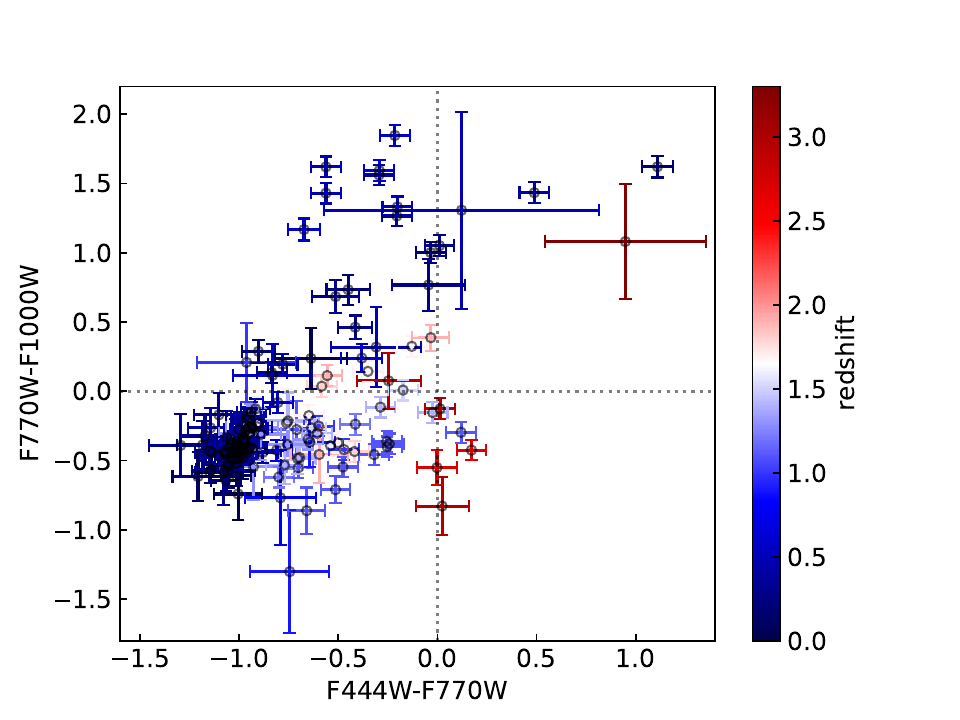}    
    \caption{Colour-colour diagram of NIRCam and MIRI bands for the matched galaxies in SMACS0723. The symbols and points are otherwise the same as in Figure~\ref{fig:mag_colors}.}
    \label{fig:colors}
\end{figure}


\begin{table}
\raggedright
\caption{$5\sigma$ depths and correction factors of magnitude-zeropoints, apertures and extinctions. }
\label{tab:corr_depth}
\begin{tabular}{lccc} 
\hline
Instrument/ Filter& Zeropoint & Aperture correction & $5\sigma$ depths\\
& AB mag & AB mag & AB mag\\
(1)&(2)&(3)&(4)\\
\hline
HST/F435W    &25.66       &-0.106      &25.14\\
HST/F606W    &26.50       &-0.095      &25.39\\
HST/F814W    &25.95       &-0.098      &25.23\\
HST/F105W    &26.27       &-0.136      &25.17\\
HST/F125W    &26.23       &-0.155      &24.87\\
HST/F140W    &26.45       &-0.164      &24.67\\
HST/F160W    &25.95       &-0.170      &25.19\\
NIRCam/F090W  &28.08      &-0.079      &27.08\\
NIRCam/F150W  &28.08      &-0.090      &26.91\\
NIRCam/F200W  &28.08      &-0.103      &26.99\\
NIRCam/F277W  &28.08      &-0.110      &27.40\\
NIRCam/F356W  &28.08      &-0.119      &27.57\\
NIRCam/F444W  &28.08      &-0.143      &27.43\\
MIRI/F770W   &28.9        &-0.202      &24.95\\
MIRI/F1000W  &28.9        &-0.326      &25.15\\
MIRI/F1500W  &28.9        &-0.369      &24.65\\
MIRI/F1800W  &28.9        &-0.421      &24.18\\
\hline
\end{tabular}
\end{table}


\section{MIRI selected galaxies}\label{sec:result}

In the following sections, we describe the main results of this paper. We outline SED fittings with and without MIRI, using \cigale\, and \eazy\, and identify the types of galaxies that are preferentially selected with MIRI included to the depths we are reaching.  Additionally, we will explore whether MIRI is capable of observing more galaxies compared to using NIRCame alone.

\subsection{Spectral energy distribution modeling}\label{sec:SED}

After generating our catalogues, we fit the spectral energy distributions of each source to derive photometric redshifts in various different ways. To calculate a preliminary photo-$z$, we fit SEDs using \cigale\, \citep{Boquien2019}. \cigale\, better constrains the fluxes on the redder bands because it includes AGN contributions and more accurate dust templates compared to \eazy\,, which we use in other EPOCHS papers \citep[e.g.,][]{Adams2023}. Here we follow the setups used by \citet{YangG2023}.

We use the standard delayed-$\tau$ `sfhdelayed' star formation history within our fitting. We set the $e$-folding time and stellar age to vary from 0.5–5 Gyr and 1–5Gyr, respectively. We use \citet{BruzualCharlot2003}(BC03) templates for the stellar population (SSP) models, assuming a \citet{Chabrier2003} initial mass function (IMF), with a solar metallicity of $Z = 0.02$. We also include within our fits the nebular module \citep{Villa2021} for emission from the HII regions, with an ionisation parameter of log\,$U = -2.0$, a gas metallicity = 0.02 and with lines width = 300.0 km/s. 

We use the `skirtor2016' module to describe the AGN component \citep{Stalevski2012,Stalevski2016}, with the fraction of AGN \fagn\, varying from 0 to 0.99 and the relative rest-frame wavelength $\lambda_{\rm AGN}$ in the range of 3–30$\mu$m. The 9.7 $\mu$m optical depths allowed in our study includes all available values 3, 5, 7, 9, and 11. We fix the AGN viewing angle to be at 70 degrees to select the obscured AGN, which is a typical value for type II AGN \citep{YangG2020,YangG2022}.

We also use the `dl2014' module developed by \citet{Draine2014} to calculate dust emission. The dust emission comprises two components: a diffused emission and a photodissociation region (PDR) emission associated with star formation. In our fitting, we allow the fraction of PDR emission ($\gamma$) to vary from 0.01 to 0.9, the minimum radiation parameter ($U{\rm min}$) to vary from 0.1, 1.0, 10, 50, and a maximum fixed value of $U{\rm max}$ = $10^7$. The mass fraction of polycyclic aromatic hydrocarbon (PAH) in total dust is the same for both components, and we set it as [0.47, 2.5, 7.32]. 
For the dust attenuation, we adopt the `dustatt' modified starburst module in \cigale\, \citep{Calzetti2000,Leitherer2002}. The colour excess is set within the range $E(B - V ) = 0 - 1$. 

In order to determine the most accurate photometric redshifts, we use the redshifting mode and a redshift grid ranging from $z = 0.0$ to 15.0, with a bin width of 0.1. We measure the properties of our sample of galaxies, including redshift, SFR, stellar mass, and \fagn\, through both traditional least-$\chi$2 analysis and different types of Bayesian approaches. The latter methods take into account the full probability density functions (PDFs), and provides more comprehensive and informative results than the least-$\chi^2$ approach \citep{Boquien2019}.

In addition, we also utilise the \eazy\,  photometric redshift code \citep{Brammer2008} to assess the accuracy of the SED fitting derived from \cigale, and \eazy, in conjunction with HST and NIRCam data. Our  \eazy\, approach involves a modified Kroupa IMF \citep{Kroupa2001} and the default templates (tweak\_fsps\_QSF\_12\_v3), which is comprised of younger stellar populations, lower metallicities, and more active star formation \citep{Larson2022}. 

\begin{figure*}
    \centering
    \includegraphics[width=18cm]{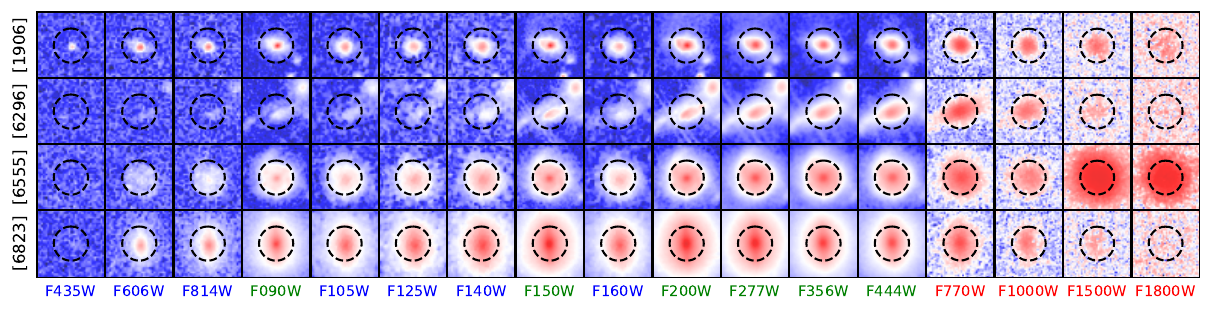}
    \caption{The different band images of a subset of the galaxies in log scale. Their IDs are labelled on the left. From left to right, the images are ACS F435W, ACS F606W, ACS F814W, NIRCam F090W, WCS3 F105W, WCS3 F125W, WCS3 F140W, NIRCam F150W, WCS3 F160W, NIRCam F200W, NIRCam F277W, NIRCam F356W, NIRCam F444W, MIRI F770W, MIRI F1000W, MIRI F1500W, MIRI F1800W. The text in blue, green, and red denotes different instruments: HST, NIRCam and MIRI, respectively. The images are 2\arcsec\,$\times$2\arcsec\, and are centred on the galaxy in each bandpass. The black circle is the aperture of 1\arcsec\,.}
    \label{fig:imshow}
\end{figure*} 

The comparison between the redshift measurements obtained from these methods reveals a high level of concordance, with deviations typically falling within 15 percent, except for a small subset of targets (8/181) fit using \eazy\, at a redshift of approximately $z\sim6$. Due to the the limited availability of dust and AGN templates at the red wavelengths, \eazy\, exhibits a less restrictive approach towards the data. It tends to primarily rely on the blue end of the data, using Lyman-break or Balmer-break techniques for redshift determination. This inclination can result in potential contamination when selecting samples with high redshifts at $z>6$. It is important to note that the occurrence of such sources is relatively scarce. Therefore, when publishing high-redshift candidates, additional stringent selection criteria need to be employed for accurate screening, a topic which is discussed in our EPOCHS paper I (Conselice in prep.). Nevertheless, our results provide strong evidence supporting the reliability and stability of our SED fitting technique when leveraging the rich photometric information provided by HST and NIRCam observations. None the less, an important conclusion from our study is that some low redshift galaxies can be mistaken for high redshift ones without the use of MIRI data.

\begin{figure*}
    \centering
    \includegraphics[width=18cm]{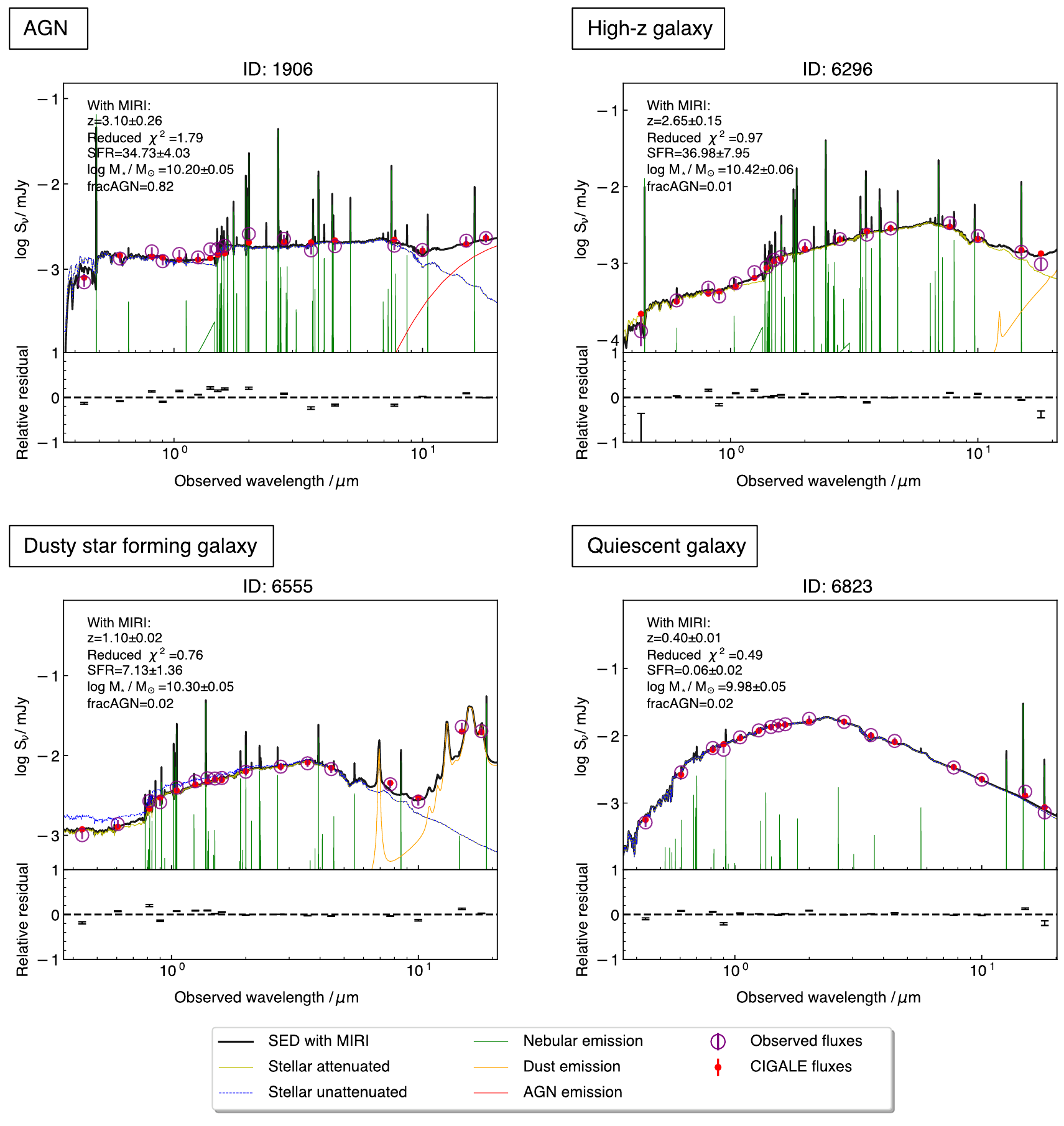}
    \caption{A subset of MIRI selected galaxies with fits done using \cigale\. Shown are systems which we classify as AGN, high-$z$ galaxies, dusty star forming galaxies, and quiescent galaxies. The black line represents the best fitting result from the \cigale\, code. The purple points represent the observed fluxes for each band; the red points represent their fitted fluxes. The yellow line represents the star formation contribution; the green line is the fitted emission line template. The red and orange lines represent the contributions of AGN and dust, respectively. The lower part of each panel is the relative residual of the fitting.}
    \label{fig:miri_selected}
\end{figure*}


\subsection{A Robust Sample of MIRI selected galaxies}\label{sec:MIRI_select}
In order to determine the physical properties of MIRI selected galaxies, we utilise the \cigale\, SED fitting approach outlined in Section~\ref{sec:SED}.  We employ a series of selection criteria described below:

\begin{enumerate}
    
    \item We require detections in both MIRI and NIRCam: $\geq 5\sigma$ detection in 2 bands in MIRI, and $\geq 5\sigma$ detections in 2 bands in NIRCam. 
    \item Removal of the sources close to the centre of the cluster, to avoid multiple imaging and excessive gravitational amplification caused by lensing.
    \item Morphology checking to exclude non-galaxy targets, e.g. hot pixels, artefacts, blended features.
    \item Matching with HST catalogue within 0.3\arcsec; for non-matched targets, we use {\tt SExtractor++} with the forced aperture to collect the flux at its position.
    \item We require $\chi^2_{red} < 6$ for best-fitting SEDs to be classed as robust.
    \item $P(z_{sec}) < 0.5 \times P(z_{phot})$ to ensure the probability of a secondary peak, if one exists, is less than 50\% of the high-$z$ solution.
\end{enumerate}

The broad emission features of PAHs in the 3-20 $\mu$m range are shifted to longer wavelengths with increasingly higher redshifts. As a result, these features are expected to dominate the flux at specific mid-infrared wavelengths, leading to significant redshift-dependent colour variations in broad-band photometry \citep{Langeroodi2023}. In Figure~\ref{fig:mag_colors}, we present the NIRCam and MIRI magnitude-colour (F444W vs. F444W-F770W) diagram for our sources, while the colour-colour (F444W-F770W vs. F770W-F1000W) diagram is shown in Figure~\ref{fig:colors}. 

As explained earlier, we determine redshifts using a Bayesian analysis based on \cigale\, fitting. In Figure~\ref{fig:mag_colors}, we observe a considerable number of cluster members that do not exhibit PAH emission and have low specific star formation rates (sSFR). Their redshifts are around $z=0.4$ and they are located at the bottom of the mag-colour plot. In Figure~\ref{fig:colors}, we find that galaxies primarily occupy the region towards the bottom left of the colour-colour diagrams, in several magnitudes of the flat-spectrum point located at position (0,0). Due to their colours, this region is likely populated by quiescent galaxies and higher redshift galaxies. 

We group our sources into the following primary categories based on the criteria above and primarily from the $\chi^2_{red}$ fits. Figure~\ref{fig:imshow} and Figure~\ref{fig:miri_selected} summarize the cutout images and SED fitting results for each category.

\begin{itemize}
  \item AGN: The emission from AGN in the MIRI bands can arise from several components. One component is the thermal emission from the dusty torus surrounding the central black hole.\citep{Fritz2006,Nenkova2008,Siebenmorgen2015} The temperature of the torus typically ranges from a few hundred to several thousand degree K, depending on the AGN's level of activity. This emission is influenced by the temperature and geometry of the torus, as well as the orientation of the system with respect to the observer. Another contribution from AGN at the MIRI bands is non-thermal emission originating from relativistic jets or outflows associated with the black hole (e.g. \citealt{Honig2017,Kakkad2023}). These high-energy particles can produce synchrotron emission in the mid-infrared regime, which can be detected by MIRI. Disentangling the AGN contribution from other sources, such as star formation, allows for a more comprehensive analysis of the galaxy's overall emission and underlying processes. We will discuss this in more detail in Section~\ref{sec:discussion_agn}.
  \item High-$z$ galaxies: With the broad wavelength coverage of the MIRI bands on JWST, several techniques can be employed to select $z>2$ galaxies. Flux dropouts or steep declines in the spectral energy distribution (SED) due to Lyman and Balmer breaks can be identified as indicators of high-redshift sources. Additionally, MIRI enables the detection of key emission lines, such as optical emission lines and \ion{O}{IV}] at 26 $\mu$m or PAH lines, which are redshifted to longer wavelengths for high-redshift sources, making them accessible in the MIRI bands. Leveraging the IR capabilities of JWST, we successfully applied the Lyman and Balmer break to select high-$z$ objects that may be undetectable or faint in blue bands like HST and F115W. MIRI photometry provides robust constraints on the SEDs, enabling precise determinations of redshift and galaxy properties. In our final catalog, we identified 46 galaxies at $z_{photo}>1$, of which 29 (63\%) have confirmed high spectroscopic-$z$ values \citep{Carnall2023,Caminha2022,Noirot2023}. For a detailed description of our extensive study on high-$z$ objects at $z > 6.5$, see our EPOCHS paper I (Conselice in prep.).
  \item Dusty star forming galaxies: MIRI offers a range of methods to search for dusty star-forming galaxies. The thermal emission from dust heated by UV/optical photons from young, massive stars can be detected using the MIRI bands. Moreover, The presence of PAH features at 6.2, 7.7, 8.6, 11.3, and 12.7 $\mu$m indicates actively star-forming galaxies, especially at high redshift (e.g. \citealt{Langeroodi2023}). Additionally, MIRI's broad wavelength coverage allows us to measure the spectral energy distribution (SED) shape and identifying characteristic features, such as the 9.7 $\mu$m silicate absorption line, providing insights into dust composition and distribution within these galaxies.\citep{Rich2023} We employ the `dl2014' module in \cigale\,, which is comprised of a diffused emission and a PDR emission associated with star formation and PAH features. This fully considers the above situation and can effectively select dusty star-forming galaxies.
  \item Quiescent galaxies: Quiescent galaxies are characterized by a low level of ongoing star formation and are typically associated with an older stellar population. These galaxies exhibit SEDs that peak at longer wavelengths, making them particularly noticeable in the MIRI bands. In the colour-colour diagram shown in Figure~\ref{fig:colors}, quiescent galaxies tend to be found within a cluster at a redshift of $z_{cl} = 0.39$ and are observed to have a colour of (F444W - F1000W)  $\sim -$0.5 mag (AB), which is consistent with the predictions of the quiescent galaxies models (Figure 1 in \citealt{Langeroodi2023}). Quiescent galaxies tend to cluster in the region towards the bottom-left of the stationary locus of the star-forming tracks. 
  
  The position of these quiescent galaxies in this region are roughly independent of redshift due to their approximately power-law SEDs. We identified all the cluster galaxies occupying the region corresponding to quiescent galaxies using spectroscopic redshifts from MUSE observations ($z = 0.387 \pm 0.02$) as reported in \citet{Caminha2022}. This is expected as various quenching mechanisms operate more efficiently in cluster environments (e.g., \citealt{Donnari2021,Kim2022}). Furthermore, in addition to the quiescent sources within the cluster, several quiescent galaxies at redshifts around $z \sim 1-2$ have been discovered within overdensities associated with a significant number of star-forming galaxies (e.g., \citealt{Noirot2023}).
\end{itemize}


We also check for sources with only MIRI detections, that are not found within NIRCam or HST observations. To ensure that we do not miss these sources, we utilised {\tt SExtractor++} and searched for detections with a 5$\sigma$ threshold or higher on at least two MIRI bands. We then measure the NIRCam and HST flux at the same positions as before, using the same aperture and mask. Interestingly, we did not find any sources that are only detected solely with MIRI, indicating that NIRCam photometry is deep enough within this field and at the MIRI depth we study to capture all the IR bright sources. The 5$\sigma$ depth of F770W and F1000W is 24.95 and 25.15 mags, which is 3 mags shallower than NIRCam F444W of 27.43 mag. This suggests that previous JWST work that relied solely on NIRCam detections is reliable in finding all galaxies to our MIRI depth.

\section{Stars, Dust, and AGN properties}\label{sec:discussion}

In this section we discuss the physical properties of our MIRI selected galaxies. We first explore their redshift, stellar mass and star formation history derived by \cigale\, fitting and then we investigate how MIRI can improve the accuracy of these measurements. Additionally, we also analyse the AGN contribution and conduct a detailed study of the median SED of the selected galaxies.

\begin{figure}
    \centering
    \includegraphics[width=8.5cm]{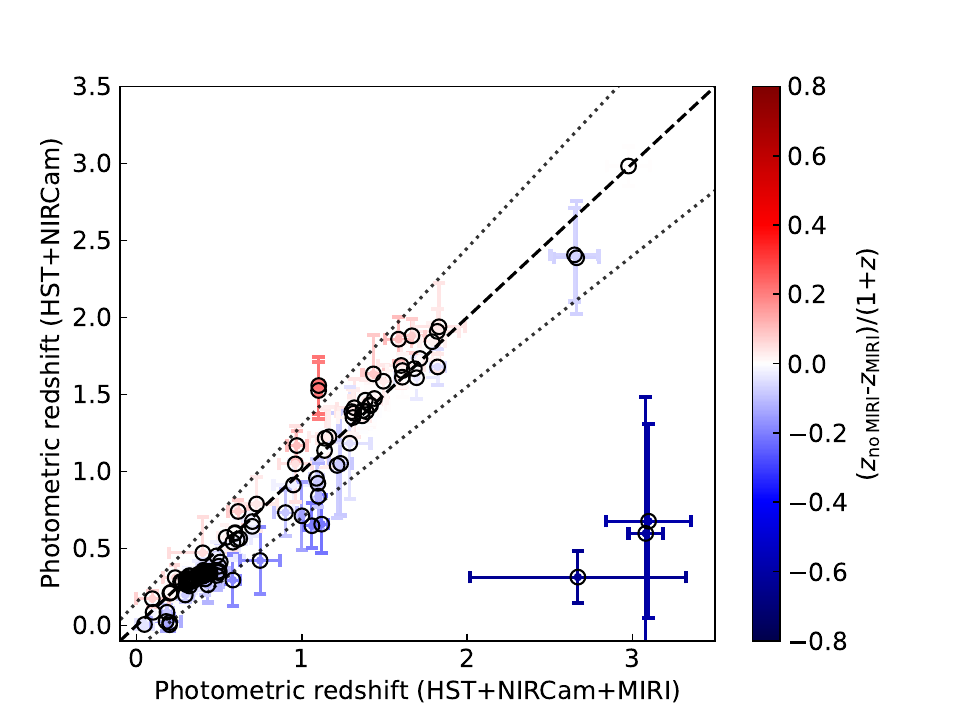}
    \caption{Comparison of the photometric redshifts derived by \cigale\, fitting with and without MIRI data. The black dashed line shows the one-to-one relation, which is the ideal 1:1 matching case of photometric vs. spectroscopic redshifts. The dotted lines show 15\,percent offsets in ($1+z$). The colour of the point represents the relative difference between the photometric redshifts of the galaxies with and without MIRI.}
    \label{fig:redshift}
\end{figure}

\begin{figure*}
 \centering
 \includegraphics[width=\textwidth]{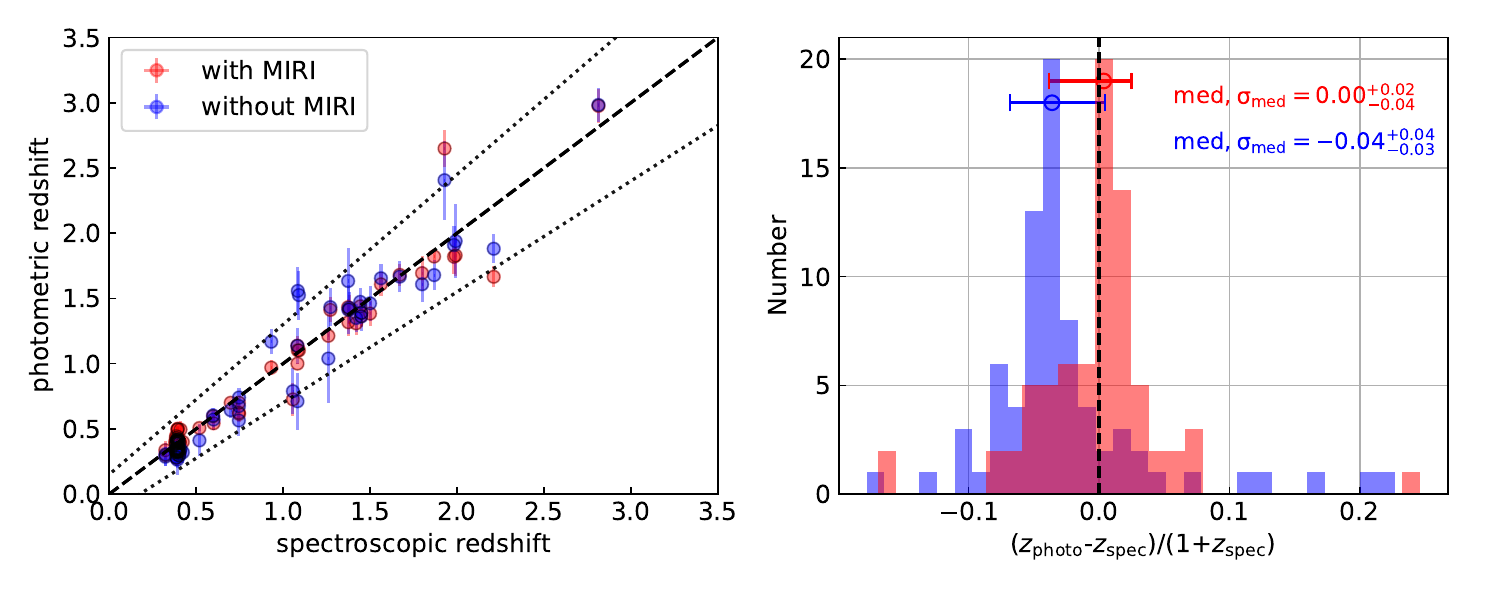}    
    \caption{Left: Diagnostic plot showing the comparison of spectroscopic redshifts with photometric redshifts for fits with and without MIRI data. The spectroscopic redshifts are from the observations of Subaru, VLT/MUSE, JWST/NIRISS and JWST/NIRSpec \citep{Carnall2023,Caminha2022,Noirot2023}. The black dashed line shows the one-to-one relation; the dotted lines show 15\,percent offsets in ($1+z$).  
    Right: the histogram of the relative difference between the photometric redshifts from our \cigale\, fits with or without MIRI and the spectroscopic redshift in $(1+z_{\rm spec})$. The labelled scatter indicates the median of the relative difference, respectively. The error bars show the range of the 16th-84th percentiles.}
    \label{fig:specz}
\end{figure*}

\subsection{The impact of MIRI on redshift measurement}\label{sec:impact_z}

Limited by the available JWST observations, most recent redshift measurement works only focus on the NIRCam analysis (e.g., \citealt{Adams2023,Bouwens2023,Endsley2023}). Here we test how and if MIRI improves the accuracy of redshift measurements. We use the \cigale\, code again to determine the redshift with and without MIRI data. The parameters in the fit are the same as before. The results show that the redshifts are nearly consistent, as shown in Figure~\ref{fig:redshift}. These two methods have photometric redshift solutions within 15 per cent of that with MIRI. In addition, we find \cigale\, fitting with MIRI data decreases the uncertainty of redshifts $\rm (\sigma_{MIRI}-\sigma_{no MIRI})/\sigma_{MIRI}$ by 50\%.

In Figure~\ref{fig:redshift}, there are three objects that stand out as outliers with a difference greater than $\Delta\,z>2$. When using MIRI to measure photometric redshifts, these objects are at high redshifts $z_{phot}>2.5$, whereas without MIRI, the derived redshift is $z<1.0$. The identification of good photometric redshifts relies on either the Lyman break or Balmer break. While fitting without MIRI data, the photo-$z$ code fits the gap between HST/ACS F435W and F606W as the Balmer break, thereby identifying them as being low redshift. However, fitting with MIRI data could change the measurement of redshift in two aspects. Firstly, MIRI data could improve constraints on the dust emission/attenuation at the redwards wavelength. Secondly, another factor to consider is the impact of nebular emission lines, including the PAH feature, on the flux in certain bands. This can potentially cause significant changes in the photometric redshift solutions. In such cases, the code fits the observed NIRCam/F200W excess as a Balmer break, resulting in a high-$z$ solution.

Although there are currently 17 multiband data points available in this field that effectively and accurately distinguish between high-$z$ and low-$z$ targets, it is evident that relying solely on photometry still creates significant uncertainties. Currently, 85 (50\%) of our galaxies have spectroscopic redshift information available. In Figure~\ref{fig:specz}, we present a comparison between the spectroscopic redshift and the photometric redshift with and without MIRI. The spectroscopic redshifts are measured by Suburu, VLT/MUSE, JWST/NIRISS and JWST/NIRSpec\citep{Carnall2023,Caminha2022,Noirot2023}. The photometric redshift data are almost all located within 15\% of the spectroscopic redshift. It can be seen that the photometric redshift is quite reliable to a certain extent, even when utilising only HST and NIRCam data. This is due to the fact that the Lyman break/Balmer break is the basis for the photometric redshift, which relies more heavily on data from the blue end. In contrast, an absence of HST data can cause a significant bias in the photometric redshift.

Figure~\ref{fig:specz} (right) displays the relative difference between the spectral redshifts and photometric redshifts with and without MIRI data. This reveals that median photometric redshift estimates have a scatter of $0.00_{-0.04}^{+0.02}$\,(0.1\%) and $-0.04_{-0.03}^{+0.04}$\,(4.0\%) from the spectroscopic redshift for fits with and without MIRI data, respectively. The outlier fractions, defined as the fraction of photometric redshift that disagrees with the spectroscopic redshift by more than 15\% in $(1+x)$, ($|\Delta\,z|$/(1+spec-$z$)$>$0.15), are 1\% and 5\%, respectively. Additionally, the results obtained from fitting with MIRI data show a closer alignment with the spectroscopic redshift and reduce the estimated errors on the photometric redshift by $\rm (\sigma_{X}-\sigma_{spec})/\sigma_{spec}$ of 20\%.

At present, spectroscopic observations are mostly at low redshifts. In the SMACS0723 field, only 10 sources with a redshift greater than 6.5 have been observed by NIRCam, and unfortunately, they have not been covered by MIRI observations. JWST mid-infrared and spectroscopic observations are still lacking at this stage. Upcoming follow-up studies are expected to provide more data, which will help to systematically constrain their redshifts and physical properties.

\subsection{Stellar Mass and Star Formation History}

\begin{figure*}
    \centering
    \includegraphics[width=18cm]{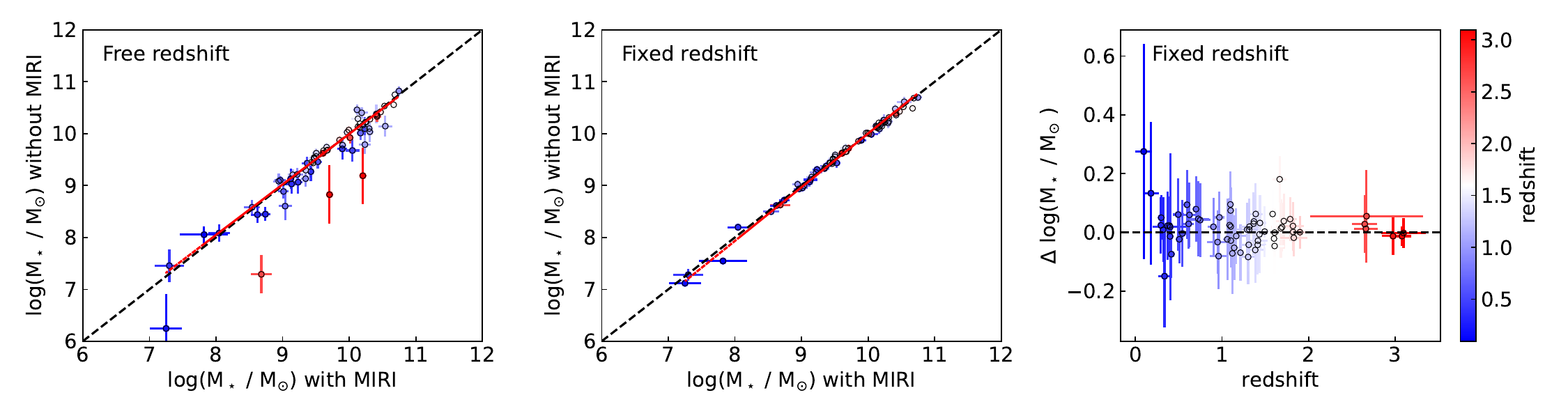}
    \includegraphics[width=18cm]{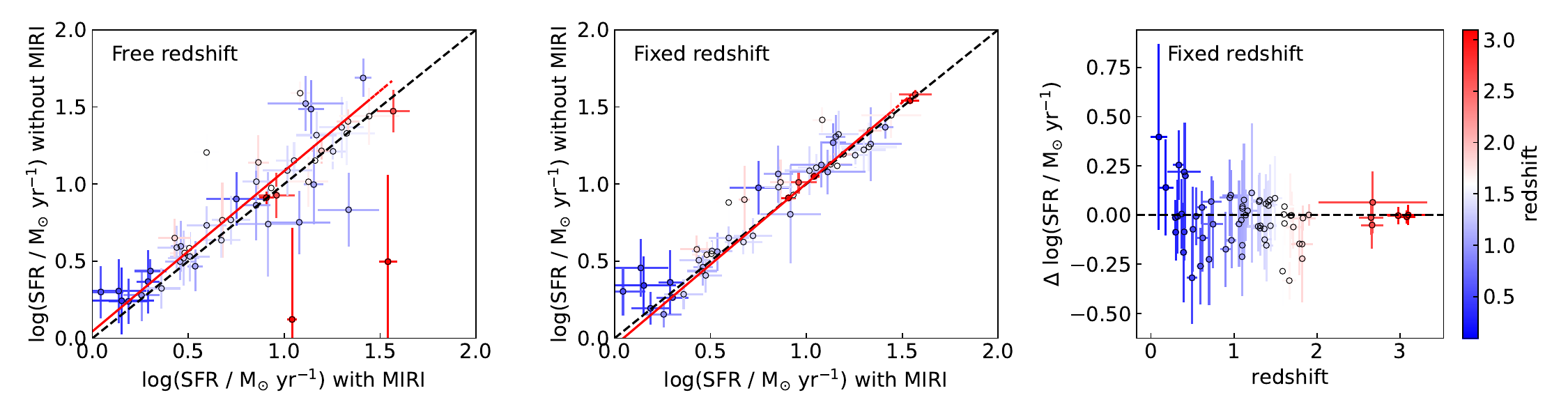}
    \caption{Comparisons between the derived stellar masses and star formation rates when including and excluding MIRI data. SFR and stellar masses are taken from the \cigale\, SED fitting, as discussed in Section~\ref{sec:SED}. The left panel shows the comparisons of stellar masses and SFR respectively, when the redshift is a free parameter. The black dashed line shows the one-to-one relation, while the red line shows the best polyfitting considering the error. In the middle and right panels, the redshifts are fixed to the values obtained from fitting the MIRI data. 
    The right panel shows the difference ($\Delta$ = X$_{\rm MIRI}$ - X$_{\rm no\,MIRI}$) as a function of redshift. 
    The colours of the points indicate the redshift. The stellar mass and SFR are not corrected for magnification, however this does not impact our results.}    \label{fig:miri}
\end{figure*}

Here we discuss the comparisons between star formation rate and masses derived when we include MIRI data and we excluded MIRI data as shown in Figure~\ref{fig:miri}. We employ the standard delayed-$\tau$ `sfhdelayed' star formation history and the bc03 stellar population module \citep{BruzualCharlot2003}, assuming using Chabrier2003 IMF \citep{Chabrier2003}. We have excluded the galaxies from our analysis, which positioned exceptionally close to the cluster's center. Thus we have not corrected the gravitational amplification for these physical parameters. In the present discussion focusing on the impact of MIRI on data fitting, the gravitational amplification does not have effect on our conclusions.

In the Figure~\ref{fig:miri} left panel, the majority of stellar mass values fall within a 15\% error range. Only a few galaxies lie away from the 1:1 line, but have a large error of $>1$ dex. The range of preferred values for stellar mass and SFR have been narrowed down with the inclusion of MIRI data. The median $\Delta M_{\star}$ error decreases 0.1 dex. This is a result of improved constraints on the dust emission and attenuation.

For the star formation rate, \cigale\, provides several SFR indicators based on different timescales: instantaneous SFR, as well as SFRs averaged over the last 10 Myrs and 100 Myrs. Generally, the SFR averaged over the last 100 Myrs is considered the most reliable indicator of the stable star-formation activity. Here we follow this custom to use the SFR averaged over the last 100 Myrs. We have excluded the quiescent galaxies with a low star formation rate of log\,sSFR\,$<-10$\,yr$^{-1}$ from this comparison. 

The SFRs derived with MIRI data are generally slightly lower by $\sim0.1$ dex. \citet{Papovich2023} also reported that adding the MIRI data could reduce SFRs for the galaxies with $\Delta$SFR of 0.15 dex at $4<z<6$ and 0.29 dex at $z>6$, matching our findings. However, for two high-$z$ objects, the log SFRs fitted with MIRI data increase by more than three times, and the error bars also significantly decrease. This is because they are identified as low-$z$ objects with a large uncertainty when we exclude MIRI data. In contrast, adding MIRI changes the best-fitting redshifts, so that they are both $z\sim3$ objects.

In the middle and right panels of Figure~\ref{fig:miri}, we re-run the \cigale\, fitting with fixed redshift values obtained from fitting with MIRI data. This was done to eliminate the influence of redshift on the results. We can see that the results show good agreement. The results indicate that the impact on the galaxy mass and SFR measurements is primarily a consequence of changing the redshift. This effect can be attributed to the additional information provided by MIRI's mid-infrared observations, allowing for a better constraint on the galaxy's redshift and, consequently, improving the accuracy of its mass and SFR determinations.

\begin{figure}
    \centering
    \includegraphics[width=8.5cm]{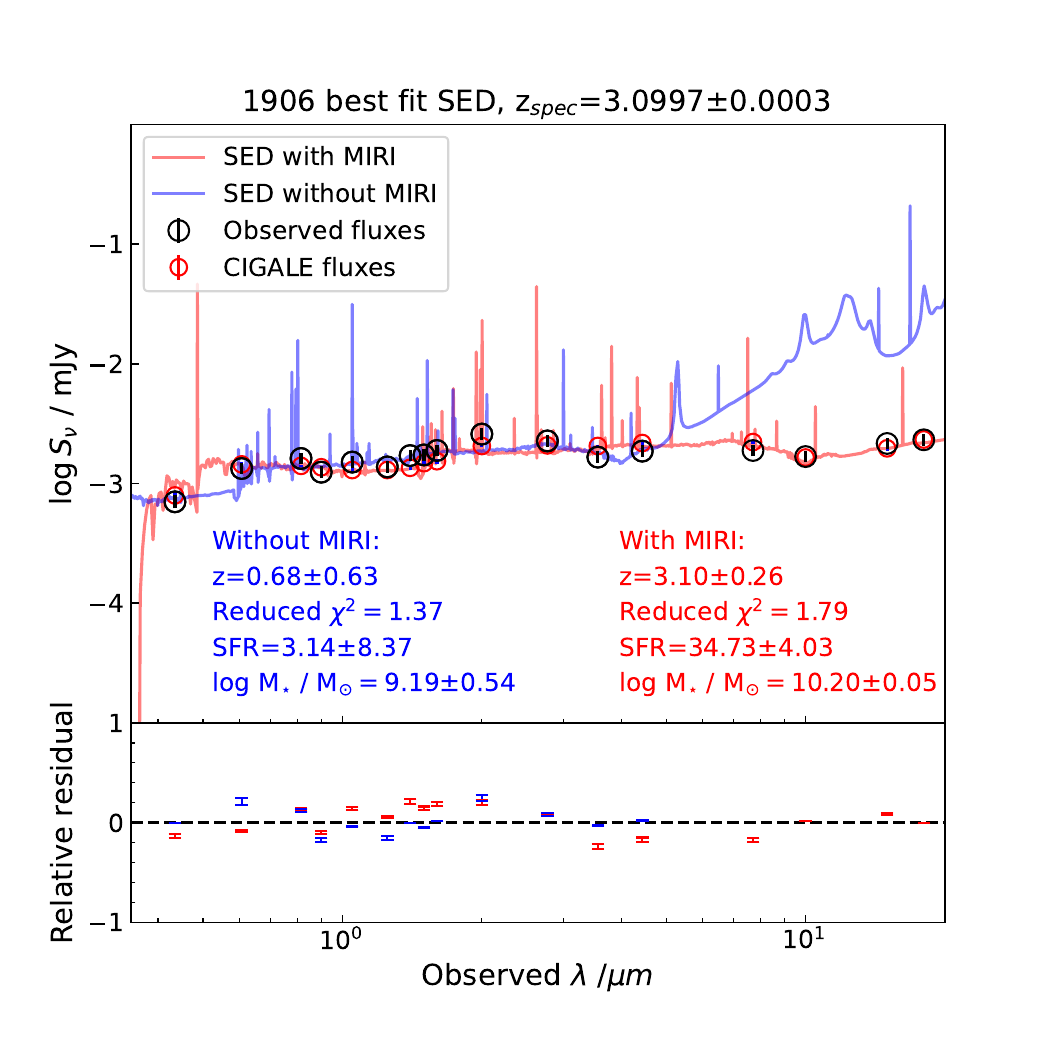}
    \caption{An example for different ways of fitting the SEDs in our sample.  We show here the \cigale\, SED fitting for Galaxy ID: 1906 with or without MIRI data. The black open circles are observed fluxes in each band; while the red ones are \cigale\, Bayesian best fitted fluxes. The red and blue lines are the best fit SEDs with and without MIRI, respectively. The bottom panel is the relative residual of the observed data points and the fitting results.  There are no blue points at long wavelengths as in this situation there is no data here.}
    \label{fig:case_z}
\end{figure}

Figure~\ref{fig:case_z} illustrates a representative example of a single galaxy fit, highlighting the significant impact of including MIRI data. The absence of MIRI data results in a loss of constraints at the red-end of the fit, leading to potential inaccuracies in various physical parameters such as redshift determinations. This emphasizes the crucial role of MIRI data in improving the accuracy and reliability of galaxy characterization and analysis.

Generally speaking, including MIRI data gives approximately similar measurements of stellar masses and SFRs to we only using NIRCam and HST. We also find that MIRI reduces the error of the stellar masses and SFRs by $\sim0.1$ dex, 
narrowing down the preferred values
of stellar population parameters. In some cases, the large differences are always caused by the redshift uncertainties.

\begin{figure*}
    \centering
    \includegraphics[width=0.49\textwidth]{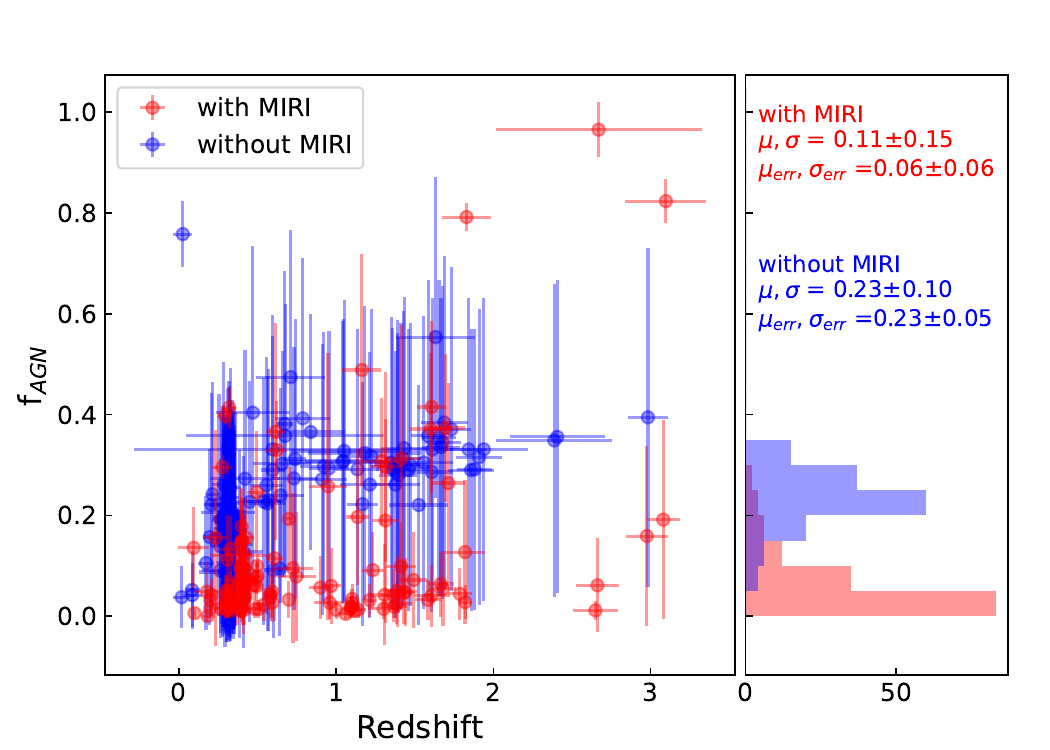}
    \includegraphics[width=0.49\textwidth]{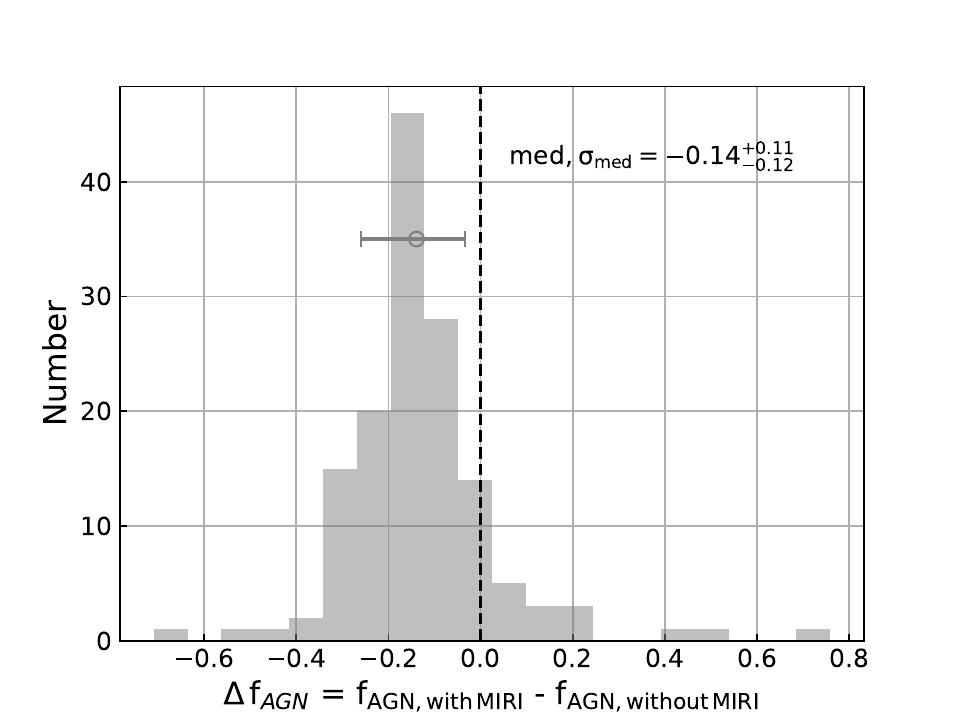}
    \caption{Left: The inferred AGN fraction(\fagn\,) as a function of redshift with and without MIRI data.  Right: the distribution of the difference \fagn\, ($\rm \Delta\,frac_{AGN}=frac_{AGN, MIRI}-frac_{AGN, no\,MIRI}$) for galaxies with and without MIRI in the fits. The median value for this difference is $-0.14 _{-0.12} ^{+\,0.11}$, similar to what is found in the MIRISIM simulation of CEERS imaging  \citet{Yang2021} who find a value  $-0.2$. The error bars show the range of the 16th-84th percentiles. 
    }
    \label{fig:fagn}
\end{figure*}

\subsection{The impact of MIRI on AGN contribution}\label{sec:discussion_agn}

We measure the contribution of AGN to our sample based on the best-fit \fagn\, parameter from \cigale\, fitting, and distinguish the galaxies between star-forming galaxies and AGNs (referred to as SFG and AGN, respectively). The dale2014 module provides a basic template from the ultraviolet to the infrared for \cigale\, fitting. The AGN fraction (\fagn\,) is defined as the ratio of the AGN luminosity to the sum of the AGN and dust luminosities \citep{Boquien2019}. It is particularly sensitive to data in the red-end at wavelengths of 3 microns to several hundred microns, where the dominant emission is primarily attributed to the AGN.   Thus, we are not using a binary approach to determine if a galaxy is all AGN or all 'stars', but we are determining from this fitting what fraction of the light emitted arises from AGN.

In Figure~\ref{fig:fagn} we conduct a test to investigate the impact of including or excluding MIRI data on the \fagn\, measurement. Our findings indicate that \fagn\, has a mean value of 0.10$\pm$0.15 in the fit that includes the MIRI data points, which is smaller than the result that does not include MIRI, where we get a fraction of 0.23$\pm$0.10. This implies that the MIRI data lower the derived fraction of the AGN and that often the contribution is higher without the use of MIRI. The median \fagn\, difference between with MIRI and without MIRI is $-0.14 _{-0.12} ^{+\,0.11}$. In \citet{Yang2021}, the MIRISIM simulation of CEERS imaging yielded a $\rm \Delta\,frac_{AGN}=(frac_{AGN, MIRI}-frac_{AGN, no\,MIRI})$ value of $\sim -0.2$ in Figure 12 Bottom panel, which aligns with our findings of $-0.14$. In addition, the inclusion of MIRI has caused a significant decrease of $\sim 0.17$ in the error of mean \fagn\,, similar to the effect on redshift and other galaxy parameters. 

However, it becomes challenging to constrain the model in the early Universe, which results in a substantial increase in the error. For instance, at $z<3$, the hot dust heated by the AGN is well tracked by the MIRI band, with a peak at 10$\mu$m in the rest frame. On the contrary, at $z>5$, the key emission from AGN-heated hot dust is shifted beyond MIRI detection ranges. The F1800W band corresponds to the rest frame wavelength of 3 microns, where the contribution of AGN has just started and is still relatively weak. This introduces significant challenges in the pursuit of identifying and investigating AGN beyond a redshift of $z>3$. We refer readers to see our other paper in this series \citep{Juodvzbalis2023}, dedicated to clarifying the complications and strategies entailed in probing AGN at $z\sim6$.

\begin{figure*}
 \includegraphics[width=0.49\textwidth]{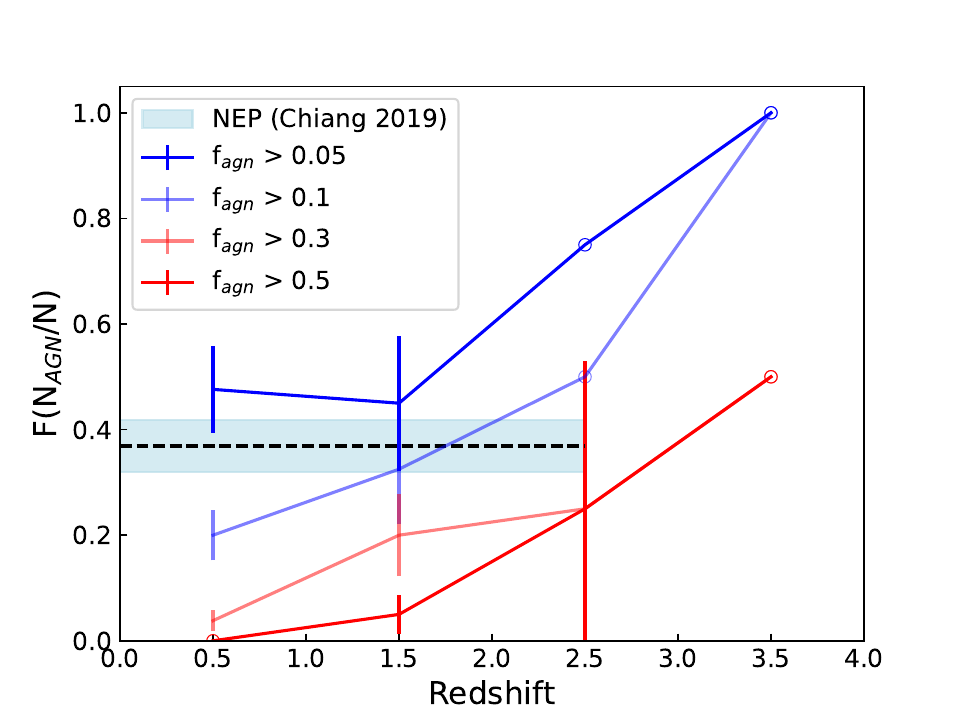}  
 \includegraphics[width=0.49\textwidth]{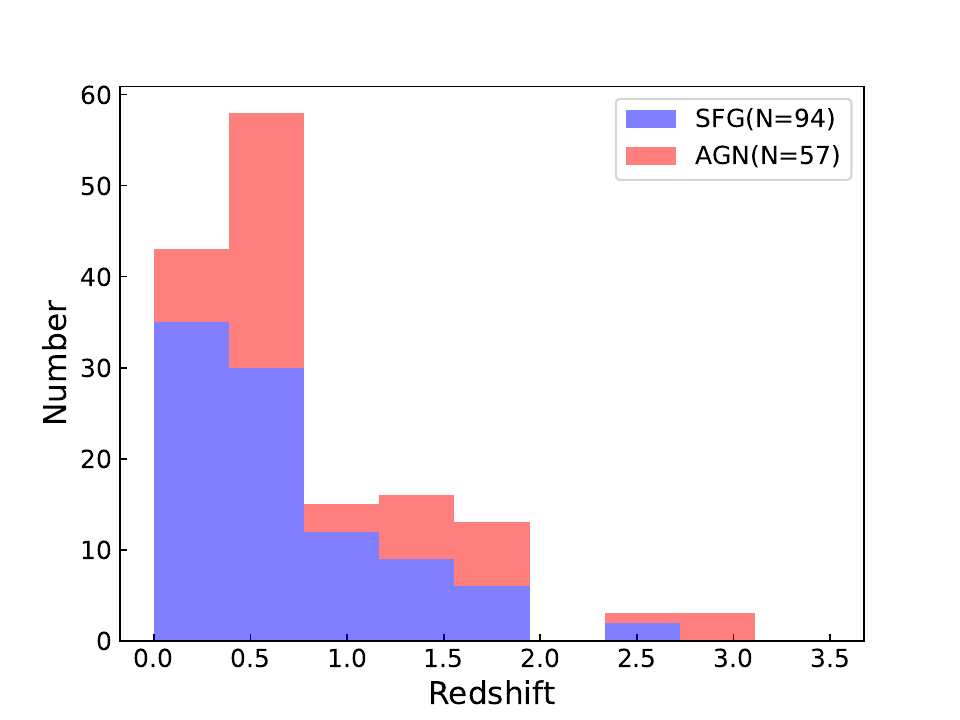}    
    \caption{Left: The proportion of AGN to the total number of galaxies as a function of redshift. We compare different \fagn\, values [0.05, 0.1, 0.3, 0.5]. We mark the points with significant uncertainty greater than 1 as open circles. With the NEP observational statistics (although these galaxies are at different redshifts and magnitudes), we conclude that a \fagn\, value of 0.1 is appropriate. The data shows that 37.7\% of the sample consists of AGN in this case. Therefore, we classify objects with \fagn\, values less than 0.1 as SFG and those above as AGN. Right: The redshift distribution of AGN and SFG is categorized based on this \fagn\,=0.1 limit.}
    \label{fig:z_distribution}
\end{figure*}

\subsection{SED analysis constraining AGN and dusty contributions}\label{sec:discussion_median_SED}

In this section, we analyse the median SEDs of AGN and SFGs using similar redshift ranges and with accurate photometric redshifts. We also investigate the effects of including MIRI data on the median SEDs.

Using \fagn\, to identify AGN is not a strict criterion, and the value we use is somewhat arbitrary. To ensure the plausibility of our results, we tested different \fagn\, values, ranging from 0.05 to 0.5, to calculate the proportion of AGN to the total number of galaxies, aiming to closely approximate the actual observed results. First, we select 151/181 best-fitting galaxies ($\chi^2 < 6$) and verify to ensure they exhibit a good fit in the red-end of the SED. As a comparison \cite{Chiang2019} used the Northern Ecliptic (NEP) wide-area catalogue who identified 6070 active galactic nuclei out of a total of 16464 IR-selected galaxies. Whilst this catalogue of galaxies is quite different from the JWST sample as the redshifts and magnitudes of sources are different,  as well as having more bands, it does show how one can use our method to find a reasonable selection for AGN.  The fitting for this NEP catalogue used \lephare\, fitting to find AGN. This NEP catalogue consists of 18 mid-infrared filters, including 9 from AKARI, 4 from WISE, and 5 from Spitzer. Our dataset exhibits a comparable redshift distribution within the range of $0<z<2.5$, close to that of the NEP sample. Using similar methods as ours, the total proportion of AGN in this NEP catalogue is 36.9$\pm$0.5\%, similar to what we find, however our systems are much fainter. Figure~\ref{fig:z_distribution} left illustrates the results, and we find that \fagn\,=0.1 is most consistent with the NEP observation statistics. In this case, the proportion of AGN (57/151) is 37.7\%.

As a result, we can conclude that if an object has a \fagn\, value of less than 0.1, we classify it as SFG; otherwise, we classify it as AGN. Using this criterion, we identified 94 SFG and 57 AGN. Figure~\ref{fig:z_distribution} right shows the photo-$z$ distributions for different types of objects. We note that slightly altering these empirical classification criteria would not significantly affect our main results.

\begin{figure*}
    \centering    \includegraphics[width=18cm]{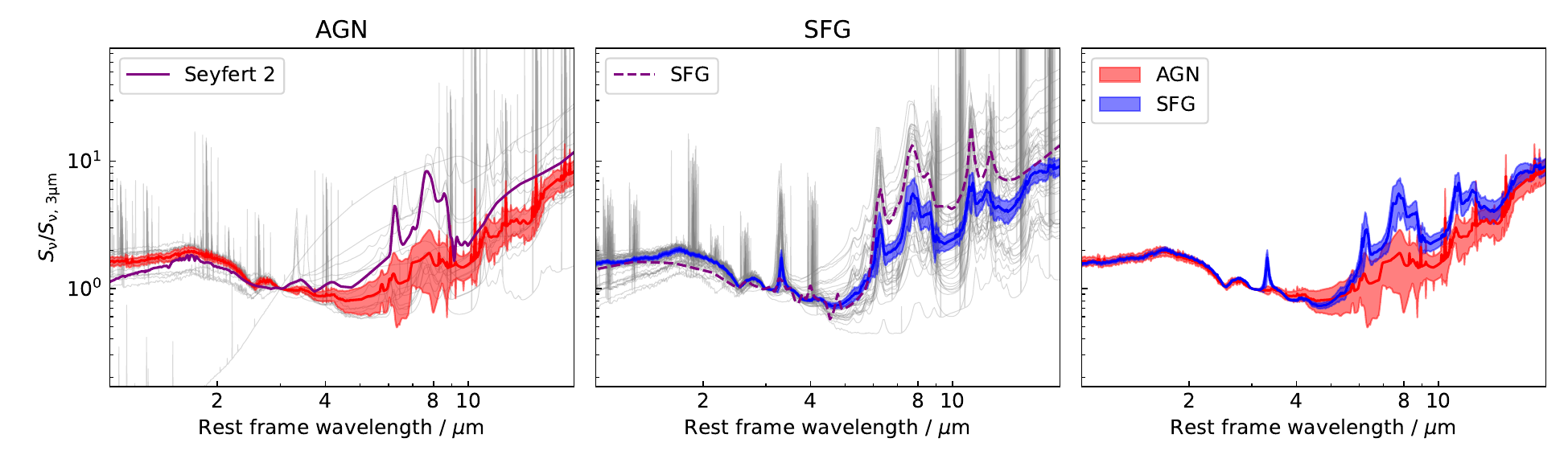} 
    \caption{The median SEDs of AGN and SFGs fitting with MIRI using \cigale. The gray lines indicate the individual robust fitting models, shifted to the rest-frame. The models are all normalized at 3 microns. The median SED and its error are obtained by sampling 5000 times using the bootstrap method. 
    The purple solid line is a Seyfert 2 galaxy template from the SWIRE Template Library{\protect\footnotemark}; the purple dashed line is the star forming galaxy template. These templates are also normalized at 3 microns.
    }
    \label{fig:agn_sed}
\end{figure*}

\footnotetext{SWIRE Template Library: \url{http://www.iasf-milano.inaf.it/~polletta/templates/swire_templates.html}}

\begin{figure}
    \includegraphics[width=8cm]{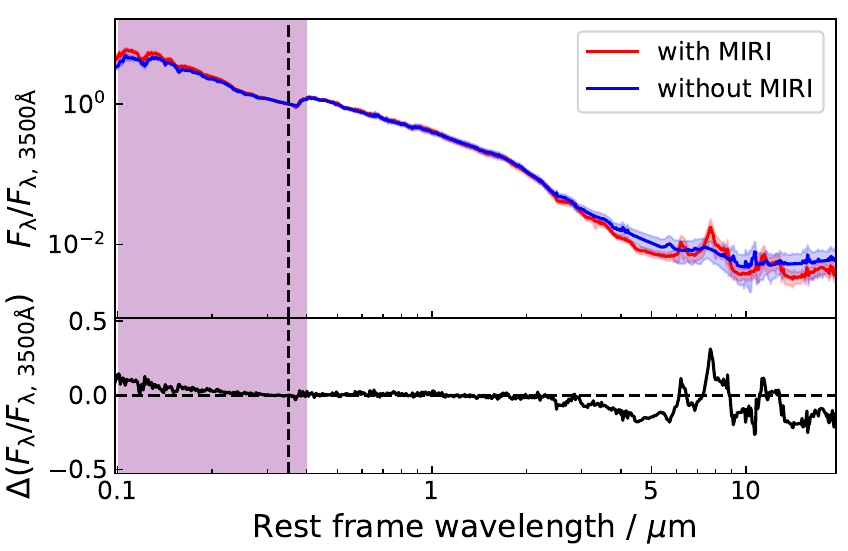} 
    \includegraphics[width=8cm]{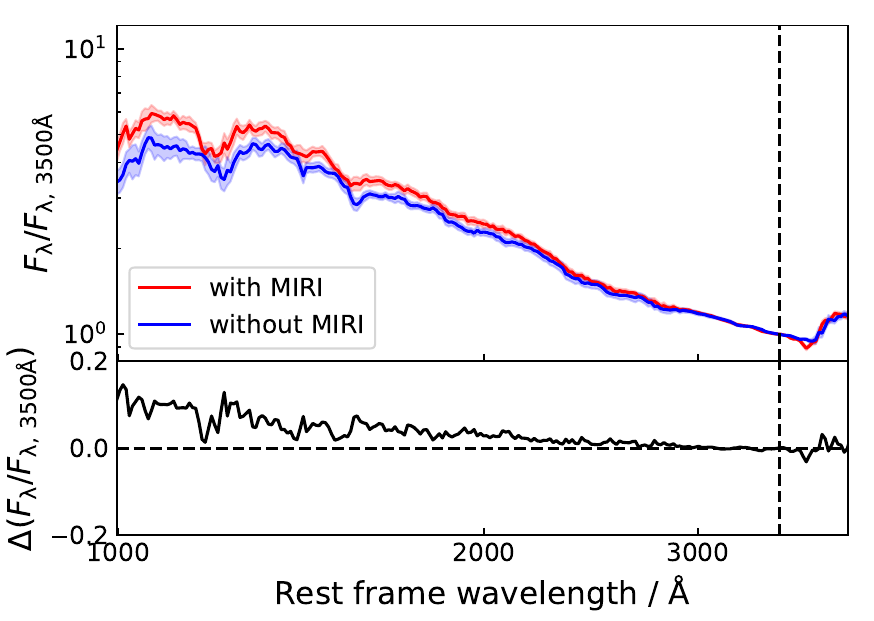}
    \caption{Comparison of median SEDs fitted when we include or exclude MIRI data. The SEDs have been normalized to 3500 \AA\,, shown as the dashed line. 
    The bottom panel is a zoom-in view in the range from 1000 to 4000 \AA\,.
    The lower part of each panel shows the difference between best fit SEDs with or without MIRI data (X$_{\rm MIRI}$-X$_{\rm no\, MIRI}$), plotted as the black line. }
    \label{fig:sed_miri}
\end{figure}

One way see how different the AGN and star forming galaxies are in our sample is to compare their SEDs. When generating the median SEDs, we first exclude the quiescent galaxies, which have no discernible PAH features ({\tt qpah}$<1$) and a lack of ongoing star formation activity, such as ID:6823 shown in Figure~\ref{fig:miri_selected}. Here {\tt qpah} is the mass fraction of the PAH \citep{Boquien2019}. Some of these galaxies may correspond to foreground cluster members.

Then, we use the photometric redshift obtained from the \cigale\, fit including MIRI to convert the best fitting models to its rest frame wavelength. We perform a linear interpolation for each model, ranging from 0.1 to 20 microns. Next, we use the bootstrap method to conduct 5000 repetitions calculating the median value and its error. Finally, we normalize the models at 3 microns, where the impact of emission lines and PAH can be avoided. We also employ a similar methodology to compute the median SED solely based on photometric data points, thereby mitigating the influence of fitting uncertainties. It is consistent in ensemble with that generated from the models. 

Figure~\ref{fig:agn_sed} shows the median SED for both the AGN and SFG objects. The grey lines indicate each individually fitted model. The SEDs are relatively constant at wavelengths below approximately $4\mu$m. But the slope of the SEDs begins to change at longer wavelengths as a result of the presence of dust and AGN. It is evident that AGN and dust greatly contribute to the red wavelengths. At redshifts of $z=0-3.5$, MIRI F770W corresponds to a rest wavelength of $2-8 \mu$m, and F1800W corresponds to $4-18 \mu$m. In this case, the MIRI data is responsible for fitting data larger than $2 \mu$m. Note that we do not differentiate between different redshift bins due to a limited number of samples, but all our sample's photometric redshifts are less than 3.5. Thus, the results are not significantly impacted by a very wide redshift distribution and range. 

We over-lay on these SEDS a moderately luminous AGN - Seyfert 2 galaxy template\footnote{SWIRE Template Library: \url{http://www.iasf-milano.inaf.it/~polletta/templates/swire_templates.html}} and a star forming galaxy template. The MIRI-selected SFGs exhibit strong dust emission and prominent PAH features. Their median SED closely resembles that of typical starburst galaxies. The median AGN SED is similar to Seyfert 2 in the ensemble sense, but has lower 6-9$\mu$m PAH emission. The 6-9$\mu$m emission primarily arises from highly vibrationally excited cations, whereas the 3.3$\mu$m, 8.6$\mu$m, and 11.3$\mu$m originate mostly from neutral PAH molecules (e.g., \citealt{Allamandola1989,Aigen2001,Draine2021}). 

The varying ratios, such as 6.2$\mu$m or 7.7$\mu$m / 11.3$\mu$m, indicate differences in the PAH ionization fraction (e.g., \citealt{Galliano2008,Rigopoulou2021}). AGN SEDs have a slightly lower average at 6.2 and 7.7$\mu$m compared to star-forming galaxies. This suggests a lower fraction of ionized PAH molecules in AGN-dominated systems from within our sample. These findings align with a PAH study on Seyfert galaxies and star-forming galaxies using Spitzer/InfraRed spectral data in \citet{GarciaBernete2022}. They imply that the nuclear molecular gas concentration in AGN centers may play a role in shielding their PAH molecules.

We emphasize that our current MIRI data points only rely on broadband photometric data. This approach may omit PAH characteristic lines, leading to inadequate fitting. To address this limitation, MIRI medium-resolution spectrometer (MRS) can provide high-resolution spectra, enabling us to determine PAH characteristic lines and mid-infrared band physical parameters more accurately.

\subsection{The impact of MIRI on median SED fitting}\label{sec:discussion_MIRI_SED}

One of the things we investigate in this subsection is the impact of MIRI data on the overall shape and form of SEDs.  What we are interested in examining is how different these SEDs would be with and without MIRI data.
Figure~\ref{fig:sed_miri} shows the median SED and the difference when fitting the data with and without MIRI. The SED difference is not noticeable at wavelengths less than 4 microns. However, at longer wavelengths, including MIRI data leads to prominent PAH features compared to the case without it (Figure~\ref{fig:sed_miri} top panel). This is because the absence of MIRI data would make it impossible to constrain the PAH emission line details in mid-infrared bands. But the dust continuum exhibits a similarity between the two cases. The \cigale\, fitting procedure guesses a relatively accurate model of dust emission, which aligns with the actual properties of the galaxies under investigation. Note the quiescent galaxies were excluded from the analysis due to their infrared SED shapes that deviate significantly from those of other galaxies.

At the rest wavelength between 4000~\AA\ and 1 micron, we find that including MIRI data in the fitting process yields a slightly steeper optical slope, though the effect is less pronounced. We also investigate the SEDs shown in Figure~\ref{fig:sed_miri} (bottom), when it comes to light which is emitted at wavelengths less than 4000~\AA\. We calculate the rest-frame UV slope ($\beta$) by fitting a power-law model of the form $f_{\lambda}\propto\lambda^{\beta}$ to the UV photometry within the range $1250\text{\AA}<\lambda_{\text{rest}}<3000\text{\AA}$ using SED fitting \citep{Bouwens2009,Finkelstein2012,Calzetti1994}. 

The best-fitted average UV slope with MIRI data is $\beta=-1.84\pm0.01$; whereas it is $\beta=-1.68\pm0.01$ without MIRI. This indicates that the MIRI selected galaxies exhibit bluer colours, lower levels of dust attenuation, and younger stellar populations. This finding is also pointed out in \citet{Papovich2023} for the CEERS field. It is important to note that the resolution of MIRI broadband photometry data points may not be sufficient to accurately identify key spectral lines, leading to inaccuracies in the existing median SED.  In the future, further research using MIRI/MRS would improve our understanding of SED in the mid-infrared band.

\section{Conclusions}\label{sec:Conclusion}

In this eighth article of the EPOCHS series, we collect data from JWST/MIRI to analyse the field SMACS0723, which is the first public release of data from this instrument from JWST. In this study, we focus on the overlapping region between the MIRI, NIRCam and HST observations, covering an area of approximately 2.3 arcmin$^2$. Within this region, we select 181 sources from a MIRI based catalogue and measure their photometric redshifts. Furthermore, we conduct an extensive investigation of various properties, including star formation activity, stellar mass, and contributions from active galactic nuclei (AGN). Our primary findings include:

\begin{itemize}
  \item We use MIRI, NIRCam, and HST data to determine these galaxies' photometric redshifts of the range of $z=0-3.5$. Furthermore, we conduct a detailed analysis of the stellar populations and the star formation and dust properties of each galaxy with and without theuse of MIRI data.
  
  \item We conduct a comparison between the photometric redshifts obtained with and without MIRI data, and cross-check them with existing spectroscopic redshifts. We find the results of the photometric redshifts are in good agreement with spectroscopic redshifts. Including MIRI data leads to an average 0.1\% difference between photometric and spectroscopic redshifts, which is 3\% lower than the difference without MIRI data. Additionally, the fitting error has also been reduced by 20\%. The redshifts of three galaxies vary by as much as $\Delta z>2$, and there are instances where high redshift galaxies would incorrectly be put at low-z without the use of MIRI data. The photometric redshifts with MIRI are highly consistent with spectroscopic redshifts, showing that the MIRI fits are better.
  
  \item We compare stellar masses and SFRs measured with and without MIRI data. Including MIRI is consistent with stellar mass measurements obtained only from HST and NIRCAM, while the SFR is slightly reduced systematically by 0.1 dex. Moreover, MIRI data also led to a decrease in both parameter errors by an average of $\sim$0.1dex.
  
  \item We select 151 the best fitting galaxies ($\chi^2 < 6$) and categorize these using the parameter \fagn\, where we consider galaxies with a value $>0.1$ AGN. Out of the total samples, 37.7\% (57/151) are found to be AGN. We determine the median values for AGN and SFG respectively. Our findings suggest that AGN and dust have a great impact on the long-wavelength flux, which is covered by the MIRI bands. Compared with the SED template, we find the SFGs match the starburst galaxy template very well. We also find that including MIRI data significantly reduces the mean value of \fagn, to 0.11$\pm$0.15, with its uncertainty also decreased of $\rm \Delta \mu_{err}=0.17$.
  
  \item We compare the median SEDs of our sample with and without MIRI data. We find that at wavelengths greater than $4\mu$m, including MIRI data reveals significant PAH features, while the dust continuum remains similar. Including MIRI data yields steeper optical and UV slopes, indicating bluer colours, lower dust attenuation, and younger stellar populations. 
  
  At present, the MIRI observations remain relatively shallow, with an average depth approximately 3 mags shallower than that of NIRCam in the SMACS0723 field. Extending the depth of MIRI observations in the future will open up a promising avenue to explore the intricacies of these galaxies in detail, and to enable the discovery of fainter and hidden galaxies. Moreover, future research utilising MIRI/MRS will improve comprehension of SEDs in the mid-infrared band and offer a more efficient approach to get redshifts and star formation rates. Through combining this with spectroscopic observations, a more detailed and nuanced illustration of the galaxies' emissions, dust properties, and other significant attributes can be achieved.  
\end{itemize}

\section*{Acknowledgements}

QL, CC, JT, and NA acknowledge support from the ERC Advanced Investigator Grant EPOCHS (788113). DA and TH acknowledge support from STFC in the form of PhD studentships.

This work is based on observations made with the NASA/ESA \textit{Hubble Space Telescope} (HST) and NASA/ESA/CSA \textit{James Webb Space Telescope} (JWST) obtained from the \texttt{Mikulski Archive for Space Telescopes} (\texttt{MAST}) at the \textit{Space Telescope Science Institute} (STScI), which is operated by the Association of Universities for Research in Astronomy, Inc., under NASA contract NAS 5-03127 for JWST, and NAS 5–26555 for HST. The observations used in this work are associated with JWST program 2736. The authors thank all involved in the construction and operations of the telescope as well as those who designed and executed these observations.

The authors thank Anthony Holloway and Sotirios Sanidas for their providing their expertise in high performance computing and other IT support throughout this work. This work makes use of {\tt astropy} \citep{Astropy2013,Astropy2018,Astropy2022}, {\tt matplotlib} \citep{Hunter2007}, {\tt reproject}, {\tt DrizzlePac} \citep{Hoffmann2021}, {\tt SciPy} \citep{2020SciPy-NMeth} and {\tt photutils} \citep{Bradley2022}.





\bibliographystyle{mnras}
\bibliography{mnras_template} 

\begin{thebibliography}{}
\makeatletter
\relax
\def\mn@urlcharsother{\let\do\@makeother \do\$\do\&\do\#\do\^\do\_\do\%\do\~}
\def\mn@doi{\begingroup\mn@urlcharsother \@ifnextchar [ {\mn@doi@}
  {\mn@doi@[]}}
\def\mn@doi@[#1]#2{\def\@tempa{#1}\ifx\@tempa\@empty \href
  {http://dx.doi.org/#2} {doi:#2}\else \href {http://dx.doi.org/#2} {#1}\fi
  \endgroup}
\def\mn@eprint#1#2{\mn@eprint@#1:#2::\@nil}
\def\mn@eprint@arXiv#1{\href {http://arxiv.org/abs/#1} {{\tt arXiv:#1}}}
\def\mn@eprint@dblp#1{\href {http://dblp.uni-trier.de/rec/bibtex/#1.xml}
  {dblp:#1}}
\def\mn@eprint@#1:#2:#3:#4\@nil{\def\@tempa {#1}\def\@tempb {#2}\def\@tempc
  {#3}\ifx \@tempc \@empty \let \@tempc \@tempb \let \@tempb \@tempa \fi \ifx
  \@tempb \@empty \def\@tempb {arXiv}\fi \@ifundefined
  {mn@eprint@\@tempb}{\@tempb:\@tempc}{\expandafter \expandafter \csname
  mn@eprint@\@tempb\endcsname \expandafter{\@tempc}}}

\bibitem[\protect\citeauthoryear{{Adams} et~al.,}{{Adams}
  et~al.}{2023}]{Adams2023}
{Adams} N.~J.,  et~al., 2023, \mn@doi [\mnras] {10.1093/mnras/stac3347}, \href
  {https://ui.adsabs.harvard.edu/abs/2023MNRAS.518.4755A} {518, 4755}

\bibitem[\protect\citeauthoryear{{Allamandola}, {Tielens}  \&
  {Barker}}{{Allamandola} et~al.}{1989}]{Allamandola1989}
{Allamandola} L.~J.,  {Tielens} A.~G.~G.~M.,   {Barker} J.~R.,  1989, \mn@doi
  [\apjs] {10.1086/191396}, \href
  {https://ui.adsabs.harvard.edu/abs/1989ApJS...71..733A} {71, 733}

\bibitem[\protect\citeauthoryear{{Asboth} et~al.,}{{Asboth}
  et~al.}{2016}]{Asboth2016}
{Asboth} V.,  et~al., 2016, \mn@doi [\mnras] {10.1093/mnras/stw1769}, \href
  {https://ui.adsabs.harvard.edu/abs/2016MNRAS.462.1989A} {462, 1989}

\bibitem[\protect\citeauthoryear{{Ashby} et~al.,}{{Ashby}
  et~al.}{2015}]{Ashby2015}
{Ashby} M.~L.~N.,  et~al., 2015, \mn@doi [\apjs] {10.1088/0067-0049/218/2/33},
  \href {https://ui.adsabs.harvard.edu/abs/2015ApJS..218...33A} {218, 33}

\bibitem[\protect\citeauthoryear{{Astropy Collaboration} et~al.,}{{Astropy
  Collaboration} et~al.}{2013}]{Astropy2013}
{Astropy Collaboration} et~al., 2013, \mn@doi [\aap]
  {10.1051/0004-6361/201322068}, \href
  {https://ui.adsabs.harvard.edu/abs/2013A&A...558A..33A} {558, A33}

\bibitem[\protect\citeauthoryear{{Astropy Collaboration} et~al.,}{{Astropy
  Collaboration} et~al.}{2018}]{Astropy2018}
{Astropy Collaboration} et~al., 2018, \mn@doi [\aj] {10.3847/1538-3881/aabc4f},
  \href {https://ui.adsabs.harvard.edu/abs/2018AJ....156..123A} {156, 123}

\bibitem[\protect\citeauthoryear{{Astropy Collaboration} et~al.,}{{Astropy
  Collaboration} et~al.}{2022}]{Astropy2022}
{Astropy Collaboration} et~al., 2022, \mn@doi [\apj]
  {10.3847/1538-4357/ac7c74}, \href
  {https://ui.adsabs.harvard.edu/abs/2022ApJ...935..167A} {935, 167}

\bibitem[\protect\citeauthoryear{{Bertin} \& {Arnouts}}{{Bertin} \&
  {Arnouts}}{1996}]{Bertin1996}
{Bertin} E.,  {Arnouts} S.,  1996, \mn@doi [\aaps] {10.1051/aas:1996164}, \href
  {https://ui.adsabs.harvard.edu/abs/1996A&AS..117..393B} {117, 393}

\bibitem[\protect\citeauthoryear{{Bertin}, {Schefer}, {Apostolakos},
  {{\'A}lvarez-Ayll{\'o}n}, {Dubath}  \& {K{\"u}mmel}}{{Bertin}
  et~al.}{2022}]{Bertin2022}
{Bertin} E.,  {Schefer} M.,  {Apostolakos} N.,  {{\'A}lvarez-Ayll{\'o}n} A.,
  {Dubath} P.,   {K{\"u}mmel} M.,  2022, {SourceXtractor++: Extracts sources
  from astronomical images}, Astrophysics Source Code Library, record
  ascl:2212.018 (\mn@eprint {ascl} {2212.018})

\bibitem[\protect\citeauthoryear{{Boquien}, {Burgarella}, {Roehlly}, {Buat},
  {Ciesla}, {Corre}, {Inoue}  \& {Salas}}{{Boquien} et~al.}{2019}]{Boquien2019}
{Boquien} M.,  {Burgarella} D.,  {Roehlly} Y.,  {Buat} V.,  {Ciesla} L.,
  {Corre} D.,  {Inoue} A.~K.,   {Salas} H.,  2019, \mn@doi [\aap]
  {10.1051/0004-6361/201834156}, \href
  {https://ui.adsabs.harvard.edu/abs/2019A&A...622A.103B} {622, A103}

\bibitem[\protect\citeauthoryear{{Bouwens} et~al.,}{{Bouwens}
  et~al.}{2009}]{Bouwens2009}
{Bouwens} R.~J.,  et~al., 2009, \mn@doi [\apj] {10.1088/0004-637X/705/1/936},
  \href {https://ui.adsabs.harvard.edu/abs/2009ApJ...705..936B} {705, 936}

\bibitem[\protect\citeauthoryear{{Bouwens} et~al.,}{{Bouwens}
  et~al.}{2023}]{Bouwens2023}
{Bouwens} R.~J.,  et~al., 2023, \mn@doi [\mnras] {10.1093/mnras/stad1145},
  \href {https://ui.adsabs.harvard.edu/abs/2023MNRAS.523.1036B} {523, 1036}

\bibitem[\protect\citeauthoryear{{Bradley} et~al.,}{{Bradley}
  et~al.}{2022}]{Bradley2022}
{Bradley} L.,  et~al., 2022, {astropy/photutils: 1.5.0}, Zenodo,
  \mn@doi{10.5281/zenodo.6825092}

\bibitem[\protect\citeauthoryear{{Brammer}, {van Dokkum}  \& {Coppi}}{{Brammer}
  et~al.}{2008}]{Brammer2008}
{Brammer} G.~B.,  {van Dokkum} P.~G.,   {Coppi} P.,  2008, \mn@doi [\apj]
  {10.1086/591786}, \href
  {https://ui.adsabs.harvard.edu/abs/2008ApJ...686.1503B} {686, 1503}

\bibitem[\protect\citeauthoryear{{Bruzual} \& {Charlot}}{{Bruzual} \&
  {Charlot}}{2003}]{BruzualCharlot2003}
{Bruzual} G.,  {Charlot} S.,  2003, \mn@doi [\mnras]
  {10.1046/j.1365-8711.2003.06897.x}, \href
  {https://ui.adsabs.harvard.edu/abs/2003MNRAS.344.1000B} {344, 1000}

\bibitem[\protect\citeauthoryear{{Calzetti}, {Kinney}  \&
  {Storchi-Bergmann}}{{Calzetti} et~al.}{1994}]{Calzetti1994}
{Calzetti} D.,  {Kinney} A.~L.,   {Storchi-Bergmann} T.,  1994, \mn@doi [\apj]
  {10.1086/174346}, \href
  {https://ui.adsabs.harvard.edu/abs/1994ApJ...429..582C} {429, 582}

\bibitem[\protect\citeauthoryear{{Calzetti}, {Armus}, {Bohlin}, {Kinney},
  {Koornneef}  \& {Storchi-Bergmann}}{{Calzetti} et~al.}{2000}]{Calzetti2000}
{Calzetti} D.,  {Armus} L.,  {Bohlin} R.~C.,  {Kinney} A.~L.,  {Koornneef} J.,
   {Storchi-Bergmann} T.,  2000, \mn@doi [\apj] {10.1086/308692}, \href
  {https://ui.adsabs.harvard.edu/abs/2000ApJ...533..682C} {533, 682}

\bibitem[\protect\citeauthoryear{{Caminha}, {Suyu}, {Mercurio}, {Brammer},
  {Bergamini}, {Acebron}  \& {Vanzella}}{{Caminha} et~al.}{2022}]{Caminha2022}
{Caminha} G.~B.,  {Suyu} S.~H.,  {Mercurio} A.,  {Brammer} G.,  {Bergamini} P.,
   {Acebron} A.,   {Vanzella} E.,  2022, \mn@doi [\aap]
  {10.1051/0004-6361/202244517}, \href
  {https://ui.adsabs.harvard.edu/abs/2022A&A...666L...9C} {666, L9}

\bibitem[\protect\citeauthoryear{{Carnall} et~al.,}{{Carnall}
  et~al.}{2023}]{Carnall2023}
{Carnall} A.~C.,  et~al., 2023, \mn@doi [\mnras] {10.1093/mnrasl/slac136},
  \href {https://ui.adsabs.harvard.edu/abs/2023MNRAS.518L..45C} {518, L45}

\bibitem[\protect\citeauthoryear{{Castellano} et~al.,}{{Castellano}
  et~al.}{2022}]{Castellano2022}
{Castellano} M.,  et~al., 2022, arXiv e-prints, \href
  {https://ui.adsabs.harvard.edu/abs/2022arXiv220709436C} {p. arXiv:2207.09436}

\bibitem[\protect\citeauthoryear{{Chabrier}}{{Chabrier}}{2003}]{Chabrier2003}
{Chabrier} G.,  2003, \mn@doi [\pasp] {10.1086/376392}, \href
  {https://ui.adsabs.harvard.edu/abs/2003PASP..115..763C} {115, 763}

\bibitem[\protect\citeauthoryear{{Chiang}, {Goto}, {Hashimoto}, {Kim},
  {Matsuhara}  \& {Oi}}{{Chiang} et~al.}{2019}]{Chiang2019}
{Chiang} C.-Y.,  {Goto} T.,  {Hashimoto} T.,  {Kim} S.~J.,  {Matsuhara} H.,
  {Oi} N.,  2019, \mn@doi [\pasj] {10.1093/pasj/psz012}, \href
  {https://ui.adsabs.harvard.edu/abs/2019PASJ...71...31C} {71, 31}

\bibitem[\protect\citeauthoryear{{Coe} et~al.,}{{Coe} et~al.}{2019}]{Coe2019}
{Coe} D.,  et~al., 2019, \mn@doi [\apj] {10.3847/1538-4357/ab412b}, \href
  {https://ui.adsabs.harvard.edu/abs/2019ApJ...884...85C} {884, 85}

\bibitem[\protect\citeauthoryear{{Donnari}, {Pillepich}, {Nelson}, {Marinacci},
  {Vogelsberger}  \& {Hernquist}}{{Donnari} et~al.}{2021}]{Donnari2021}
{Donnari} M.,  {Pillepich} A.,  {Nelson} D.,  {Marinacci} F.,  {Vogelsberger}
  M.,   {Hernquist} L.,  2021, \mn@doi [\mnras] {10.1093/mnras/stab1950}, \href
  {https://ui.adsabs.harvard.edu/abs/2021MNRAS.506.4760D} {506, 4760}

\bibitem[\protect\citeauthoryear{{Draine} et~al.,}{{Draine}
  et~al.}{2014}]{Draine2014}
{Draine} B.~T.,  et~al., 2014, \mn@doi [\apj] {10.1088/0004-637X/780/2/172},
  \href {https://ui.adsabs.harvard.edu/abs/2014ApJ...780..172D} {780, 172}

\bibitem[\protect\citeauthoryear{{Draine}, {Li}, {Hensley}, {Hunt}, {Sandstrom}
   \& {Smith}}{{Draine} et~al.}{2021}]{Draine2021}
{Draine} B.~T.,  {Li} A.,  {Hensley} B.~S.,  {Hunt} L.~K.,  {Sandstrom} K.,
  {Smith} J. D.~T.,  2021, \mn@doi [\apj] {10.3847/1538-4357/abff51}, \href
  {https://ui.adsabs.harvard.edu/abs/2021ApJ...917....3D} {917, 3}

\bibitem[\protect\citeauthoryear{{Ebeling}, {Edge}, {Mantz}, {Barrett},
  {Henry}, {Ma}  \& {van Speybroeck}}{{Ebeling} et~al.}{2010}]{Ebeling2010}
{Ebeling} H.,  {Edge} A.~C.,  {Mantz} A.,  {Barrett} E.,  {Henry} J.~P.,  {Ma}
  C.~J.,   {van Speybroeck} L.,  2010, \mn@doi [\mnras]
  {10.1111/j.1365-2966.2010.16920.x}, \href
  {https://ui.adsabs.harvard.edu/abs/2010MNRAS.407...83E} {407, 83}

\bibitem[\protect\citeauthoryear{{Endsley}, {Stark}, {Whitler}, {Topping},
  {Chen}, {Plat}, {Chisholm}  \& {Charlot}}{{Endsley}
  et~al.}{2023}]{Endsley2023}
{Endsley} R.,  {Stark} D.~P.,  {Whitler} L.,  {Topping} M.~W.,  {Chen} Z.,
  {Plat} A.,  {Chisholm} J.,   {Charlot} S.,  2023, \mn@doi [\mnras]
  {10.1093/mnras/stad1919}, \href
  {https://ui.adsabs.harvard.edu/abs/2023MNRAS.tmp.1872E} {}

\bibitem[\protect\citeauthoryear{{Ferreira} et~al.,}{{Ferreira}
  et~al.}{2022}]{Ferreira2022}
{Ferreira} L.,  et~al., 2022, \mn@doi [\apjl] {10.3847/2041-8213/ac947c}, \href
  {https://ui.adsabs.harvard.edu/abs/2022ApJ...938L...2F} {938, L2}

\bibitem[\protect\citeauthoryear{{Finkelstein} et~al.,}{{Finkelstein}
  et~al.}{2012}]{Finkelstein2012}
{Finkelstein} S.~L.,  et~al., 2012, \mn@doi [\apj]
  {10.1088/0004-637X/756/2/164}, \href
  {https://ui.adsabs.harvard.edu/abs/2012ApJ...756..164F} {756, 164}

\bibitem[\protect\citeauthoryear{{Fritz}, {Franceschini}  \&
  {Hatziminaoglou}}{{Fritz} et~al.}{2006}]{Fritz2006}
{Fritz} J.,  {Franceschini} A.,   {Hatziminaoglou} E.,  2006, \mn@doi [\mnras]
  {10.1111/j.1365-2966.2006.09866.x}, \href
  {https://ui.adsabs.harvard.edu/abs/2006MNRAS.366..767F} {366, 767}

\bibitem[\protect\citeauthoryear{{Fudamoto} et~al.,}{{Fudamoto}
  et~al.}{2017}]{Fudamoto2017}
{Fudamoto} Y.,  et~al., 2017, \mn@doi [\mnras] {10.1093/mnras/stx1956}, \href
  {https://ui.adsabs.harvard.edu/abs/2017MNRAS.472.2028F} {472, 2028}

\bibitem[\protect\citeauthoryear{{Galliano}, {Madden}, {Tielens}, {Peeters}  \&
  {Jones}}{{Galliano} et~al.}{2008}]{Galliano2008}
{Galliano} F.,  {Madden} S.~C.,  {Tielens} A. G.~G.~M.,  {Peeters} E.,
  {Jones} A.~P.,  2008, \mn@doi [\apj] {10.1086/587051}, \href
  {https://ui.adsabs.harvard.edu/abs/2008ApJ...679..310G} {679, 310}

\bibitem[\protect\citeauthoryear{{Garc{\'\i}a-Bernete}, {Rigopoulou},
  {Alonso-Herrero}, {Pereira-Santaella}, {Roche}  \&
  {Kerkeni}}{{Garc{\'\i}a-Bernete} et~al.}{2022}]{GarciaBernete2022}
{Garc{\'\i}a-Bernete} I.,  {Rigopoulou} D.,  {Alonso-Herrero} A.,
  {Pereira-Santaella} M.,  {Roche} P.~F.,   {Kerkeni} B.,  2022, \mn@doi
  [\mnras] {10.1093/mnras/stab3127}, \href
  {https://ui.adsabs.harvard.edu/abs/2022MNRAS.509.4256G} {509, 4256}

\bibitem[\protect\citeauthoryear{{Golubchik}, {Furtak}, {Meena}  \&
  {Zitrin}}{{Golubchik} et~al.}{2022}]{Golubchik2022}
{Golubchik} M.,  {Furtak} L.~J.,  {Meena} A.~K.,   {Zitrin} A.,  2022, \mn@doi
  [\apj] {10.3847/1538-4357/ac8ff1}, \href
  {https://ui.adsabs.harvard.edu/abs/2022ApJ...938...14G} {938, 14}

\bibitem[\protect\citeauthoryear{{Harikane} et~al.,}{{Harikane}
  et~al.}{2022}]{Harikane2022}
{Harikane} Y.,  et~al., 2022, \mn@doi [\apj] {10.3847/1538-4357/ac53a9}, \href
  {https://ui.adsabs.harvard.edu/abs/2022ApJ...929....1H} {929, 1}

\bibitem[\protect\citeauthoryear{{Hoffmann}, {Mack}, {Avila}, {Martlin},
  {Cohen}  \& {Bajaj}}{{Hoffmann} et~al.}{2021}]{Hoffmann2021}
{Hoffmann} S.~L.,  {Mack} J.,  {Avila} R.,  {Martlin} C.,  {Cohen} Y.,
  {Bajaj} V.,  2021, in American Astronomical Society Meeting Abstracts. p.
  216.02

\bibitem[\protect\citeauthoryear{{H{\"o}nig} \& {Kishimoto}}{{H{\"o}nig} \&
  {Kishimoto}}{2017}]{Honig2017}
{H{\"o}nig} S.~F.,  {Kishimoto} M.,  2017, \mn@doi [\apjl]
  {10.3847/2041-8213/aa6838}, \href
  {https://ui.adsabs.harvard.edu/abs/2017ApJ...838L..20H} {838, L20}

\bibitem[\protect\citeauthoryear{Hunter}{Hunter}{2007}]{Hunter2007}
Hunter J.~D.,  2007, \mn@doi [Computing in Science \& Engineering]
  {10.1109/MCSE.2007.55}, 9, 90

\bibitem[\protect\citeauthoryear{{Juod{\v{z}}balis} et~al.,}{{Juod{\v{z}}balis}
  et~al.}{2023}]{Juodvzbalis2023}
{Juod{\v{z}}balis} I.,  et~al., 2023, \mn@doi [\mnras]
  {10.1093/mnras/stad2396}, \href
  {https://ui.adsabs.harvard.edu/abs/2023MNRAS.525.1353J} {525, 1353}

\bibitem[\protect\citeauthoryear{{Kakkad}, {Stalevski}, {Kishimoto},
  {Kne{\v{z}}evi{\'c}}, {Asmus}  \& {Vogt}}{{Kakkad} et~al.}{2023}]{Kakkad2023}
{Kakkad} D.,  {Stalevski} M.,  {Kishimoto} M.,  {Kne{\v{z}}evi{\'c}} S.,
  {Asmus} D.,   {Vogt} F.~P.~A.,  2023, \mn@doi [\mnras]
  {10.1093/mnras/stac3827}, \href
  {https://ui.adsabs.harvard.edu/abs/2023MNRAS.519.5324K} {519, 5324}

\bibitem[\protect\citeauthoryear{{Kim} et~al.,}{{Kim} et~al.}{2022}]{Kim2022}
{Kim} K.~J.,  et~al., 2022, \mn@doi [arXiv e-prints]
  {10.48550/arXiv.2207.12491}, \href
  {https://ui.adsabs.harvard.edu/abs/2022arXiv220712491K} {p. arXiv:2207.12491}

\bibitem[\protect\citeauthoryear{{Kroupa}}{{Kroupa}}{2001}]{Kroupa2001}
{Kroupa} P.,  2001, \mn@doi [\mnras] {10.1046/j.1365-8711.2001.04022.x}, \href
  {https://ui.adsabs.harvard.edu/abs/2001MNRAS.322..231K} {322, 231}

\bibitem[\protect\citeauthoryear{{Lagattuta} et~al.,}{{Lagattuta}
  et~al.}{2022}]{Lagattuta2022}
{Lagattuta} D.~J.,  et~al., 2022, \mn@doi [\mnras] {10.1093/mnras/stac418},
  \href {https://ui.adsabs.harvard.edu/abs/2022MNRAS.514..497L} {514, 497}

\bibitem[\protect\citeauthoryear{{Langeroodi} \& {Hjorth}}{{Langeroodi} \&
  {Hjorth}}{2023}]{Langeroodi2023}
{Langeroodi} D.,  {Hjorth} J.,  2023, \mn@doi [\apjl]
  {10.3847/2041-8213/acc1e0}, \href
  {https://ui.adsabs.harvard.edu/abs/2023ApJ...946L..40L} {946, L40}

\bibitem[\protect\citeauthoryear{{Larson} et~al.,}{{Larson}
  et~al.}{2022}]{Larson2022}
{Larson} R.~L.,  et~al., 2022, \mn@doi [arXiv e-prints]
  {10.48550/arXiv.2211.10035}, \href
  {https://ui.adsabs.harvard.edu/abs/2022arXiv221110035L} {p. arXiv:2211.10035}

\bibitem[\protect\citeauthoryear{{Leitherer}, {Li}, {Calzetti}  \&
  {Heckman}}{{Leitherer} et~al.}{2002}]{Leitherer2002}
{Leitherer} C.,  {Li} I.~H.,  {Calzetti} D.,   {Heckman} T.~M.,  2002, \mn@doi
  [\apjs] {10.1086/342486}, \href
  {https://ui.adsabs.harvard.edu/abs/2002ApJS..140..303L} {140, 303}

\bibitem[\protect\citeauthoryear{{Li} \& {Draine}}{{Li} \&
  {Draine}}{2001}]{Aigen2001}
{Li} A.,  {Draine} B.~T.,  2001, \mn@doi [\apj] {10.1086/323147}, \href
  {https://ui.adsabs.harvard.edu/abs/2001ApJ...554..778L} {554, 778}

\bibitem[\protect\citeauthoryear{{Mahler} et~al.,}{{Mahler}
  et~al.}{2023}]{Mahler2023}
{Mahler} G.,  et~al., 2023, \mn@doi [\apj] {10.3847/1538-4357/acaea9}, \href
  {https://ui.adsabs.harvard.edu/abs/2023ApJ...945...49M} {945, 49}

\bibitem[\protect\citeauthoryear{{Medezinski} et~al.,}{{Medezinski}
  et~al.}{2007}]{2007ApJ...663..717M}
{Medezinski} E.,  et~al., 2007, \mn@doi [\apj] {10.1086/518638}, \href
  {https://ui.adsabs.harvard.edu/abs/2007ApJ...663..717M} {663, 717}

\bibitem[\protect\citeauthoryear{{Menzel} et~al.,}{{Menzel}
  et~al.}{2023}]{Menzel2023}
{Menzel} M.,  et~al., 2023, \mn@doi [\pasp] {10.1088/1538-3873/acbb9f}, \href
  {https://ui.adsabs.harvard.edu/abs/2023PASP..135e8002M} {135, 058002}

\bibitem[\protect\citeauthoryear{{Naidu} et~al.,}{{Naidu}
  et~al.}{2022}]{Naidu2022}
{Naidu} R.~P.,  et~al., 2022, arXiv e-prints, \href
  {https://ui.adsabs.harvard.edu/abs/2022arXiv220709434N} {p. arXiv:2207.09434}

\bibitem[\protect\citeauthoryear{{Nayyeri} et~al.,}{{Nayyeri}
  et~al.}{2018}]{Nayyeri2018}
{Nayyeri} H.,  et~al., 2018, \mn@doi [\apjs] {10.3847/1538-4365/aaa07e}, \href
  {https://ui.adsabs.harvard.edu/abs/2018ApJS..234...38N} {234, 38}

\bibitem[\protect\citeauthoryear{{Nenkova}, {Sirocky}, {Nikutta}, {Ivezi{\'c}}
  \& {Elitzur}}{{Nenkova} et~al.}{2008}]{Nenkova2008}
{Nenkova} M.,  {Sirocky} M.~M.,  {Nikutta} R.,  {Ivezi{\'c}} {\v{Z}}.,
  {Elitzur} M.,  2008, \mn@doi [\apj] {10.1086/590483}, \href
  {https://ui.adsabs.harvard.edu/abs/2008ApJ...685..160N} {685, 160}

\bibitem[\protect\citeauthoryear{{Noirot} et~al.,}{{Noirot}
  et~al.}{2023}]{Noirot2023}
{Noirot} G.,  et~al., 2023, \mn@doi [\mnras] {10.1093/mnras/stad1019}, \href
  {https://ui.adsabs.harvard.edu/abs/2023MNRAS.tmp.1025N} {}

\bibitem[\protect\citeauthoryear{{Oke} \& {Gunn}}{{Oke} \&
  {Gunn}}{1983}]{Oke1983}
{Oke} J.~B.,  {Gunn} J.~E.,  1983, \mn@doi [\apj] {10.1086/160817}, \href
  {https://ui.adsabs.harvard.edu/abs/1983ApJ...266..713O} {266, 713}

\bibitem[\protect\citeauthoryear{{Papovich} et~al.,}{{Papovich}
  et~al.}{2023}]{Papovich2023}
{Papovich} C.,  et~al., 2023, \mn@doi [\apjl] {10.3847/2041-8213/acc948}, \href
  {https://ui.adsabs.harvard.edu/abs/2023ApJ...949L..18P} {949, L18}

\bibitem[\protect\citeauthoryear{{Pontoppidan} et~al.,}{{Pontoppidan}
  et~al.}{2022}]{Pontoppidan2022}
{Pontoppidan} K.~M.,  et~al., 2022, \mn@doi [\apjl] {10.3847/2041-8213/ac8a4e},
  \href {https://ui.adsabs.harvard.edu/abs/2022ApJ...936L..14P} {936, L14}

\bibitem[\protect\citeauthoryear{{Repp} \& {Ebeling}}{{Repp} \&
  {Ebeling}}{2018}]{Repp2018}
{Repp} A.,  {Ebeling} H.,  2018, \mn@doi [\mnras] {10.1093/mnras/sty1489},
  \href {https://ui.adsabs.harvard.edu/abs/2018MNRAS.479..844R} {479, 844}

\bibitem[\protect\citeauthoryear{{Reuter} et~al.,}{{Reuter}
  et~al.}{2020}]{Reuter2020}
{Reuter} C.,  et~al., 2020, \mn@doi [\apj] {10.3847/1538-4357/abb599}, \href
  {https://ui.adsabs.harvard.edu/abs/2020ApJ...902...78R} {902, 78}

\bibitem[\protect\citeauthoryear{{Rich} et~al.,}{{Rich}
  et~al.}{2023}]{Rich2023}
{Rich} J.,  et~al., 2023, \mn@doi [\apjl] {10.3847/2041-8213/acb2b8}, \href
  {https://ui.adsabs.harvard.edu/abs/2023ApJ...944L..50R} {944, L50}

\bibitem[\protect\citeauthoryear{{Richard} et~al.,}{{Richard}
  et~al.}{2021}]{Richard2021}
{Richard} J.,  et~al., 2021, \mn@doi [\aap] {10.1051/0004-6361/202039462},
  \href {https://ui.adsabs.harvard.edu/abs/2021A&A...646A..83R} {646, A83}

\bibitem[\protect\citeauthoryear{{Rigby} et~al.,}{{Rigby}
  et~al.}{2023}]{Rigby2023}
{Rigby} J.,  et~al., 2023, \mn@doi [\pasp] {10.1088/1538-3873/acb293}, \href
  {https://ui.adsabs.harvard.edu/abs/2023PASP..135d8001R} {135, 048001}

\bibitem[\protect\citeauthoryear{{Rigopoulou} et~al.,}{{Rigopoulou}
  et~al.}{2021}]{Rigopoulou2021}
{Rigopoulou} D.,  et~al., 2021, \mn@doi [\mnras] {10.1093/mnras/stab959}, \href
  {https://ui.adsabs.harvard.edu/abs/2021MNRAS.504.5287R} {504, 5287}

\bibitem[\protect\citeauthoryear{{Schlafly} \& {Finkbeiner}}{{Schlafly} \&
  {Finkbeiner}}{2011}]{Schlafly2011}
{Schlafly} E.~F.,  {Finkbeiner} D.~P.,  2011, \mn@doi [\apj]
  {10.1088/0004-637X/737/2/103}, \href
  {https://ui.adsabs.harvard.edu/abs/2011ApJ...737..103S} {737, 103}

\bibitem[\protect\citeauthoryear{{Siebenmorgen}, {Heymann}  \&
  {Efstathiou}}{{Siebenmorgen} et~al.}{2015}]{Siebenmorgen2015}
{Siebenmorgen} R.,  {Heymann} F.,   {Efstathiou} A.,  2015, \mn@doi [\aap]
  {10.1051/0004-6361/201526034}, \href
  {https://ui.adsabs.harvard.edu/abs/2015A&A...583A.120S} {583, A120}

\bibitem[\protect\citeauthoryear{{Stalevski}, {Fritz}, {Baes}, {Nakos}  \&
  {Popovi{\'c}}}{{Stalevski} et~al.}{2012}]{Stalevski2012}
{Stalevski} M.,  {Fritz} J.,  {Baes} M.,  {Nakos} T.,   {Popovi{\'c}}
  L.~{\v{C}}.,  2012, \mn@doi [\mnras] {10.1111/j.1365-2966.2011.19775.x},
  \href {https://ui.adsabs.harvard.edu/abs/2012MNRAS.420.2756S} {420, 2756}

\bibitem[\protect\citeauthoryear{{Stalevski}, {Ricci}, {Ueda}, {Lira}, {Fritz}
  \& {Baes}}{{Stalevski} et~al.}{2016}]{Stalevski2016}
{Stalevski} M.,  {Ricci} C.,  {Ueda} Y.,  {Lira} P.,  {Fritz} J.,   {Baes} M.,
  2016, \mn@doi [\mnras] {10.1093/mnras/stw444}, \href
  {https://ui.adsabs.harvard.edu/abs/2016MNRAS.458.2288S} {458, 2288}

\bibitem[\protect\citeauthoryear{{Timlin} et~al.,}{{Timlin}
  et~al.}{2016}]{Timlin2016}
{Timlin} J.~D.,  et~al., 2016, \mn@doi [\apjs] {10.3847/0067-0049/225/1/1},
  \href {https://ui.adsabs.harvard.edu/abs/2016ApJS..225....1T} {225, 1}

\bibitem[\protect\citeauthoryear{{Villa-V{\'e}lez}, {Buat}, {Theul{\'e}},
  {Boquien}  \& {Burgarella}}{{Villa-V{\'e}lez} et~al.}{2021}]{Villa2021}
{Villa-V{\'e}lez} J.~A.,  {Buat} V.,  {Theul{\'e}} P.,  {Boquien} M.,
  {Burgarella} D.,  2021, \mn@doi [\aap] {10.1051/0004-6361/202140890}, \href
  {https://ui.adsabs.harvard.edu/abs/2021A&A...654A.153V} {654, A153}

\bibitem[\protect\citeauthoryear{Virtanen et~al.,}{Virtanen
  et~al.}{2020}]{2020SciPy-NMeth}
Virtanen P.,  et~al., 2020, \mn@doi [Nature Methods]
  {10.1038/s41592-019-0686-2}, \href {https://rdcu.be/b08Wh} {17, 261}

\bibitem[\protect\citeauthoryear{{Weedman} et~al.,}{{Weedman}
  et~al.}{2006}]{Weedman2006}
{Weedman} D.,  et~al., 2006, \mn@doi [\apj] {10.1086/508647}, \href
  {https://ui.adsabs.harvard.edu/abs/2006ApJ...653..101W} {653, 101}

\bibitem[\protect\citeauthoryear{{Wright} et~al.,}{{Wright}
  et~al.}{2023}]{Wright2023}
{Wright} G.~S.,  et~al., 2023, \mn@doi [\pasp] {10.1088/1538-3873/acbe66},
  \href {https://ui.adsabs.harvard.edu/abs/2023PASP..135d8003W} {135, 048003}

\bibitem[\protect\citeauthoryear{{Yan}, {Ma}, {Ling}, {Cheng}, {Huang}  \&
  {Zitrin}}{{Yan} et~al.}{2022}]{Yan2022}
{Yan} H.,  {Ma} Z.,  {Ling} C.,  {Cheng} C.,  {Huang} J.-s.,   {Zitrin} A.,
  2022, arXiv e-prints, \href
  {https://ui.adsabs.harvard.edu/abs/2022arXiv220711558Y} {p. arXiv:2207.11558}

\bibitem[\protect\citeauthoryear{{Yang} et~al.,}{{Yang}
  et~al.}{2020}]{YangG2020}
{Yang} G.,  et~al., 2020, \mn@doi [\mnras] {10.1093/mnras/stz3001}, \href
  {https://ui.adsabs.harvard.edu/abs/2020MNRAS.491..740Y} {491, 740}

\bibitem[\protect\citeauthoryear{{Yang} et~al.,}{{Yang}
  et~al.}{2021}]{Yang2021}
{Yang} G.,  et~al., 2021, \mn@doi [\apj] {10.3847/1538-4357/abd6c1}, \href
  {https://ui.adsabs.harvard.edu/abs/2021ApJ...908..144Y} {908, 144}

\bibitem[\protect\citeauthoryear{{Yang} et~al.,}{{Yang}
  et~al.}{2022}]{YangG2022}
{Yang} G.,  et~al., 2022, \mn@doi [\apj] {10.3847/1538-4357/ac4971}, \href
  {https://ui.adsabs.harvard.edu/abs/2022ApJ...927..192Y} {927, 192}

\bibitem[\protect\citeauthoryear{{Yang} et~al.,}{{Yang}
  et~al.}{2023}]{YangG2023}
{Yang} G.,  et~al., 2023, \mn@doi [\apjl] {10.3847/2041-8213/acd639}, \href
  {https://ui.adsabs.harvard.edu/abs/2023ApJ...950L...5Y} {950, L5}

\makeatother
\end{thebibliography}








\bsp	
\label{lastpage}
\end{document}